\documentclass[preprint,number]{elsarticle}

\usepackage{amsmath,amssymb}
\usepackage{cases}
\usepackage{wasysym}
\usepackage{textgreek}
\usepackage[version=4]{mhchem}
\usepackage{bigints}

\newcommand{\pd}[2]{\frac{\partial{#1}}{\partial{#2}}}
\newcommand{\pdd}[2]{\frac{\partial^2{#1}}{\partial{#2}^2}}
\newcommand{\od}[2]{\frac{\mathrm{d}{#1}}{\mathrm{d}{#2}}}
\newcommand{\odd}[2]{\frac{\mathrm{d}^2{#1}}{\mathrm{d}{#2}^2}}

\usepackage{lineno}

\journal{""}
\begin{document}
	\begin{frontmatter}
		\title{How grain structure evolution affects kinetics of a solid-state
		reaction: a case of interaction between iridium and zirconium carbide}
		
		\author[1]{Ya. A. Nikiforov\corref{cor1}}
		\ead{nikiforov@solid.nsc.ru}
		
		\author[2]{V. A. Danilovsky}

		\author[1]{N. I. Baklanova}
		
		\affiliation[1]{organization={Institute of Solid State Chemistry and
			Mechanochemistry SB RAS},
			addressline={18 Kutateladze st.},
			postcode={630090},
			city={Novosibirsk},
			country={Russia}}

		\affiliation[2]{organization={Sobolev Institute of Geology and Mineralogy
		SB RAS},
	       addressline={3/1 Koptyug ave.},
	       postcode={630090},
	       city={Novosibirsk},
	       country={Russia}}
		
		\cortext[cor1]{Corresponding author}
		
		\begin{abstract}
			This work investigates the solid-state reaction between iridium and zirconium carbide,
			resulting in the formation of carbon and \ce{ZrIr_3}---an intermetallic compound of great
			interest for modern high-temperature materials science. We have found a transition of kinetic
			regimes in this reaction: from linear kinetics (when the chemical reaction is a limiting stage)
			at 1500 and 1550\textdegree C to `non-parabolic kinetics' at 1600\textdegree C. Non-parabolic
			kinetics is characterized by thickness of a product layer being proportional to a power of
			time less than 1/2. The nature of non-parabolic kinetics was still an open question,
			which motivated us to develop a model of this kinetic regime. The proposed model accounts
			for the grain growth in the	product phase and how it leads to the time dependence of the
			interdiffusion coefficient. We have obtained a complete analytic solution for this model
			and an equation that connects the grain-growth exponent and the power-law exponent of
			non-parabolic kinetics. The measurements of the thickness of the product layer and the
			average grain size of the intermetallic phase confirm the results of the theoretical solution.
		\end{abstract}
		
	

	\end{frontmatter}

	\section{Introduction}
	Recently, iridium was proposed as a principal component for
slowly oxidating protective systems \cite{Wu2017_1,Wu2017_2}. Indeed, it has high melting point 
(2446\textdegree C \cite{Arblaster1995}) combined with low oxidation rate, 
as well as low oxygen and carbon permeability \cite{Wu2017_1,Criscione1964},
which makes iridium an excellent candidate for these purposes. Practical application
of non-alloyed iridium coatings
turned to be impossible due to recrystallization of Ir during thermal cycling 
that produced pores within coatings \cite{Wu2017_2,Mumtaz1995}, i.e. 
fast-diffusion pathways. This fact did not discourage researchers, and now new
approaches of alloying iridium or using multilayered coatings based on
combination of iridium with tantalum, hafnium, and tungsten or their
carbides/borides are explored 
\cite{Wu2011,Zhu2013,Zhu2014,Chen2014,Zhu2017,Zhang2020,Zhang2022,Li2024}.
Components of such systems interact with each other, so a successful production
and application of the coatings depends on the understanding of these interactions
and resulting time-evolution of the systems.

This work is an investigation of kinetics of a reaction between iridium and 
zirconium carbide, one of the reaction of iridium with refractory carbides. It was
shown that the reaction occurs at temperatures at least as high as 
1000\textdegree C according to equation: \ce{ZrC + 3Ir -> ZrIr_3 + C} 
\cite{Nikiforov2024}. The kinetics of this and similar reactions, though, is still
mainly not investigated. The only relevant work, known to authors, is as study 
of a reaction between Ir and HfC by Kwon \cite{Kwon1989}. Kwon investigates
formation of \ce{HfIr_3}+C layer in the reaction between Ir and HfC at temperatures
1900--2200\textdegree C and calculates interdiffusion coefficients and activation
energy of interdiffusion. Additionally, recent works on reactions between silicon
carbide and transition metals \cite{Park1999,Bhanumurthy2001,Demkowicz2008,Lopez2010,
Gentile2015,Golosov2023} are of some interest.
Similarly to the ZrC--Ir system considered here, a metal and a carbide react in those
systems, with formation of intermetallic compound(s) and carbon. It was shown,
that a number of complex phenomena occur in course of reaction, although not all
of them should be present in the case of ZrC--Ir. For example, a formation of 
periodic patterns, a common feature of reactions with SiC, was not observed in the
case of Ir--HfC system. Nonetheless, an awareness of possible processes and 
comparison between different similar systems will be useful in interpretation of data.

We use reaction couples as a method of kinetics investigation. In this method two 
planar surfaces of the reactants are made into contact, heat treated at a given
temperature, while the reaction products form between the reactants as a layer of
nearly uniform width. Rate of solid state reaction can be limited by either 
rate of a reaction at interface or by transport process within bulk of products
\cite{Harrison1969}. Our aim is to describe kinetics of the reaction and its
dependence on temperature. During the investigation we discovered a transition
between different kinetic regimes. So it was crucial to analyze these regimes
and reasons why reaction proceeds according to these regimes and why the 
transition occurs.

For reaction couples, three kinetic regimes were observed earlier in literature 
\cite{Suzuki2005,Xu2006,Ren2013,Gueydan2014,Vianco1994,Bader1995,Mita2005,
Sakama2009,Li2010,Mirjalili2013,Lis2014,Zhang2019,Oh2024}, each being a case of 
reaction-limed or transport-limited reaction rate. First one is the interface
reaction controlled regime, with the product layer thickness being linearly
proportional to time ($\ell\propto t$) at long reaction times \cite{Murray1988}.
Second regime is the diffusion controlled regime, with the product layer thickness
being proportional to the square root of time ($\ell\propto\sqrt{t}$) \cite{Wagner1969,Lichtner1986}.
Third regime is non-parabolic, or power-law kinetics, with layer thickness
being proportional to a power of time less than 0.5 ($\ell\propto t^{1/n}$ with 
$n>2$). It has been observed experimentally in some systems during last two decades
\cite{Vianco1994,Bader1995,Mita2005,Sakama2009,Li2010,Mirjalili2013,Lis2014,Zhang2019,Oh2024},
although this type of kinetics is less common than previous two. This regime is 
thought to be a type of transport process limiting, but the diffusion is not 
uniform as in the second (diffusion-controlled) regime. First was suggested a 
model of variable diffusion coefficient due to a region of enhanced diffusion
near the reaction front \cite{Erickson1994}. This diffusion enhancement was ascribed
to the microstructural intricacies of a reaction zone and to a high concentration
of inequilibrium defects. Later, it was suggested that grain growth in the product
phase is responsible for the power-law kinetics \cite{Bader1995,Schaefer1998,Ghosh2000}.
These works showed that it is the most plausible explanation of non-parabolic 
kinetics, but the problem of an accurate generalization of this model remained.
This model was developed for thin one-grain-thick intermetallics layers, and the
rate exponent ($1/n$) in this model depends not only on grain growth exponent, 
but also on relation between layer thickness and grain size. These assumptions
limit the application of the model to thin layers; meanwhile, non-parabolic 
kinetics was observed in thick layers \cite{Mirjalili2013}, too. Two
recent works \cite{Wang2015,Xu2018} examined the effect of grain growth on 
solid-state reaction kinetics numerically and showed a good agreement with the
experiment. The numerical solution showed the validity of the proposed model and
made it possible to examine cases of different ratios of bulk and grain-boundary
diffusion coefficients. But, as with any numerical solution, it cannot provide 
an explicit functional dependence. So, a question remains: is the power law 
really a consequence of the grain growth within the product layer and can be 
derived analytically within this model, or is it only an approximation to the 
actual kinetic law? Thus, we aim to derive an explicit analytical solution for
this model of non-parabolic growth with a minimal number of approximations. 

\section{Theory}

\begin{figure}[h]
	 \centering
	 \includegraphics{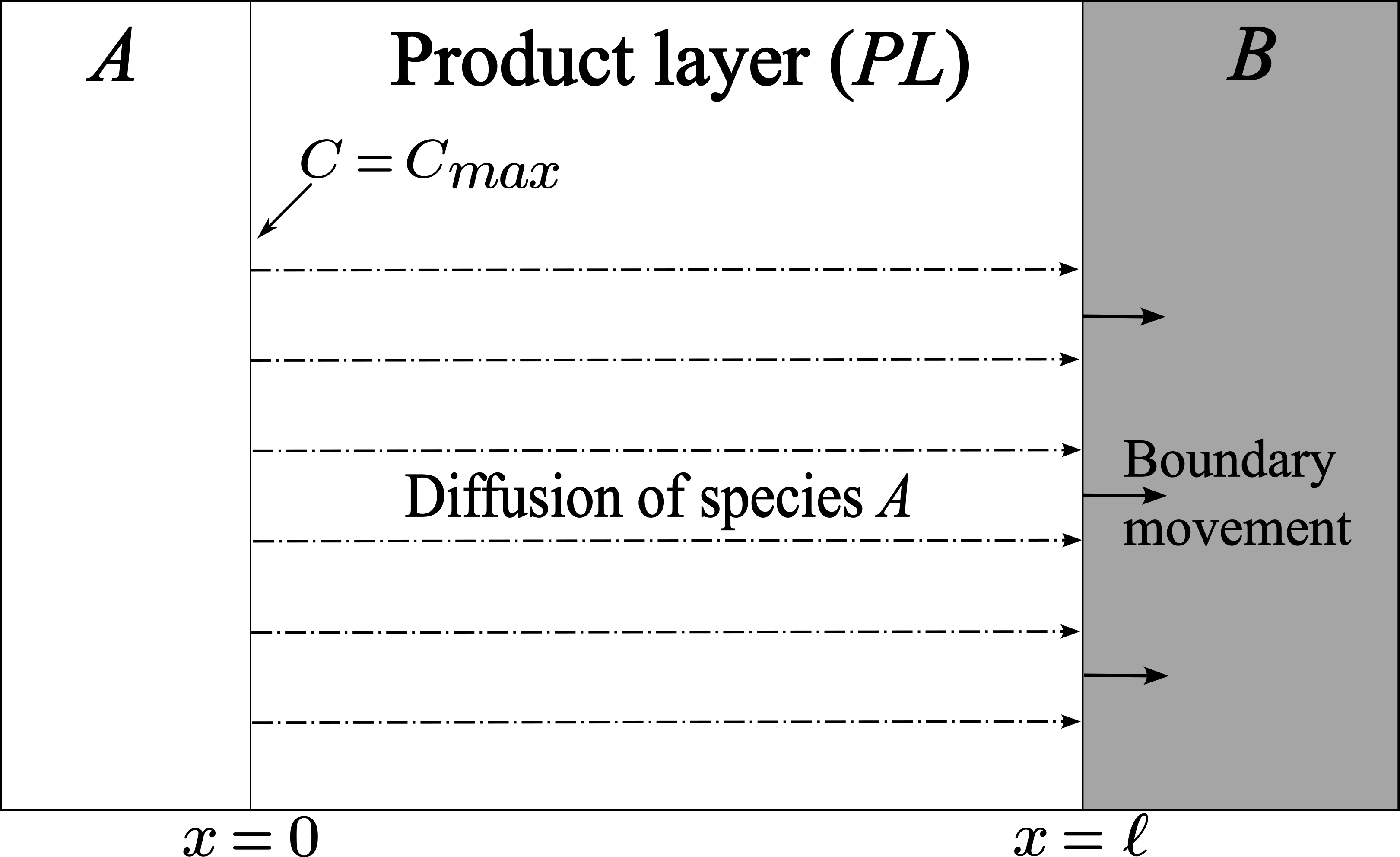}
	 \caption{Scheme of the reaction couple $A + B \rightarrow PL$. Concentration of \textit{A}
     at interface \textit{A}/\textit{PL} is $C_{max}$, meanwhile concentration 
	 at the interface \textit{PL}/\textit{B} may or may not be fixed depending
	 on the particular kinetics regime.}
	 \label{fig:rc_setup}
 \end{figure}

To analyze the kinetics of a solid-state reaction in reaction couples, we describe 
mathematical models of these. Consider a situation schematically illustrated in
Fig.~\ref{fig:rc_setup}. Two phases---\textit{A} and \textit{B}---react to form
a product layer: $A + B \rightarrow PL$. The reaction takes place
at an interface between \textit{PL} and \textit{B} (reactive boundary), while 
the atoms of \textit{A} diffuse through the product layer \textit{PL}. Atoms of
\textit{A} cross the boundary \textit{A}/\textit{PL} without the formation of
a new volume of the product so that no reaction occurs, while concentration at this
boundary remains constant. In this model, \textit{PL} does not need to be a
single phase---it is only necessary that a single interdiffusion coefficient 
describes the diffusion through PL. Governing equations are:
\begin{gather}
	\pd{C}{t} = \pd{}{x}\left(D\pd{C}{x}\right), \quad 0\leq x\leq \ell(t) \label{eq:fick_eq_d_const}\\
	C(0, t) = C_{max} \label{eq:first_bc}\\
	\frac{1}{\tilde{V}_A}\od{\ell}{t} = k \cdot (C - C_{eq}) 
 = -D \left( \pd{C}{x} \right) \bigg|_{x=\ell} \label{eq:stef_bc}\\
 \ell(0) = 0 \label{eq:width_ic}
\end{gather}

Eq.~\ref{eq:fick_eq_d_const} is the second Fick's law with spatial domain
from interface \textit{A/PL} ($x=0$) to the reactive boundary \textit{PL/B}
($x = \ell(t)$, i.e. width of \textit{PL}); $D$ is an interdiffusion 
coefficient. Eq.~\ref{eq:first_bc} is a boundary condition on interface 
\textit{A/PL}. Eq.~\ref{eq:stef_bc} is a condition on the movement of boundary
\textit{PL/B}: the first equal sign connects it with the rate of reaction at the 
interface \textit{PL/B}, and the second equal sign connects it with the diffusional
flux of atoms \textit{A} coming from the bulk of \textit{PL} towards the reactive
boundary. $\tilde{V}_A$ is a volume of \textit{PL} that forms when one atom
\textit{A} reacts with \textit{B}. $C_{eq}$ is an equilibrium concentration of
\textit{A} in \textit{PL} that is in contact with \textit{B}. Eq.~\ref{eq:width_ic}
is an initial condition on the width of \textit{PL}.
 
\subsection{Models of kinetics with constant interdiffusion coefficient}
Commonly, an interdiffusion coefficient in Eqs.~\ref{eq:fick_eq_d_const}--\ref{eq:width_ic}
is assumed to be constant. With this assumption, two limiting cases are possible,
depending on the ratio $D/k$: either the interface reaction or the transport of
\textit{A} through \textit{PL} determines the overall reaction rate. These
limiting cases simplify the equation so that analytical solutions can be 
conveniently obtained. The derivations of these solutions can be found in 
\cite{Murray1988} and \cite{Lichtner1986}. Because the notation in this work
differs from \cite{Murray1988} and \cite{Lichtner1986}, we also give derivations
in \ref{ap:sols}.

For a case of interface reaction control, diffusion of \textit{A} is much faster
than interface reaction, and a simplification is that the concentration profile
$C(x, t)$ is assumed to be linear. The solution \cite{Murray1988} is:
 \begin{numcases}{}
	 \ell(t) = \frac{D}{k} \left( \sqrt{1 + 2(C_{max} - C_{eq})\tilde{V}_A \frac{k^2}{D} t} - 1 \right)
	 \label{eq:react_control}\\
	C(x, t) = C_{max} -
	\frac{(C_{max} - C_{eq})x}{\sqrt{1 + 2(C_{max} - C_{eq})\tilde{V}_A \frac{k^2}{D} t}} 
	\label{eq:conc_prof_react}
 \end{numcases}

 In a case of diffusion control, the interface reaction occurs much faster than
 diffusion within \textit{PL}. It leads to a local equilibrium on the reactive 
 boundary so that $C(\ell, t) = C_{eq}$, and movement of the reactive boundary is 
 determined by the concentration gradient. The solution \cite{Lichtner1986} is:
 \begin{numcases}{}
	\ell(t) = 2 \gamma\sqrt{Dt} \label{eq:par_gr}\\
	C(x, t) = C_{max} - \frac{C_{max} - C_{eq}}
		{\operatorname{erf}(\gamma)}\operatorname{erf}\left(\frac{x}{2 \sqrt{Dt}}\right)
		\label{eq:c_ord_diff}
		\\
	\frac{(C_{max} - C_{eq})\tilde{V}_A}{\sqrt{\pi}} = \gamma \operatorname{erf}(\gamma)
	\exp(\gamma^2) \label{eq:gamma_cond_par}
 \end{numcases}

Eqs.~\ref{eq:react_control} and \ref{eq:par_gr} are the kinetic laws that
can be used to analyze growth of product layer in solid-state reaction.

\subsection{Grain-growth accompanied kinetics}
\label{ssec:gg_kinetics}

Here, we consider a situation when the interdiffusion coefficient is not constant
and is affected by the grain growth in the product phase. We study this case
because it was earlier proposed \cite{Bader1995,Schaefer1998,Ghosh2000} that
the grain growth is a reason for non-parabolic kinetics, which is described with
the power law $\ell\propto t^{1/n}$ rather than with Eqs. 5 and 7. Our goal here is
to obtain an analytical solution for this situation with a minimal number of
assumptions. An alternative derivation, more general than here, can be found in
\ref{ap:compl}.

We treat this situation as a case of diffusion control, that is the velocity
of the movement of reactive boundary is determined by a diffusion flux. Governing
equations for this case are:
\begin{gather}
	\pd{C}{t} = \pd{}{x}\left(D(x,t)\pd{C}{x}\right), \quad 0\leq x\leq \ell(t) \label{eq:fick_eq_d_var}\\
	C(0, t) = C_{max}, \quad C(\ell, t) = C_{eq} \label{eq:diff_bound_cond}\\
	\frac{1}{\tilde{V}_A}\od{\ell}{t}  = -D \left( \pd{C}{x} \right)
	\bigg|_{x=\ell} \label{eq:stef_bc_32}\\
 \ell(0) = 0 \label{eq:diff_init_cond}
\end{gather}

To solve these equations, we need at first to obtain expression for the 
interdiffusion coefficient. Expression for interdiffusion coefficient \cite{Mishin1997}in 
terms of a bulk diffusion coefficient, $D_B$, and a grain-boundary diffusion
coefficient, $D_{GB}$, is:
\begin{equation*}
	D = g D_{GB} +  (1-g)D_B
\end{equation*}
where, $g$ is an effective volume fraction available for grain boundary diffusion 
and is given by the equation:
\begin{equation}
	g = \frac{q\delta}{L}
	\label{eq:grain_frac}
\end{equation}
where $L$ is an average grain size, $\delta$ is a grain boundary width and $q$ 
is a numerical factor that depends on the shape of grains.

The first approximation we use is that the bulk diffusion is negligible in comparison
with grain boundary diffusion, $gD_{GB} \gg (1-g)D_B$. With that, the effective 
interdiffusion coefficient is:
\begin{equation}
	D = gD_{GB}
	\label{eq:diff_coef_appr}
\end{equation}

Next, we need to account for the grain growth, which obeys the power law 
\cite{Abbaschian2008,Molodov2013}:
\begin{equation}
	L^m - L_0^m = k(t - t_0)
	\label{eq:grain_growth}
\end{equation}
where $L$ is an average grain size at time $t$, $L_0$ is an average grain size
at initial time $t_0$, $k$ is a constant of grain growth. 

The second approximation is that the initial grain size is negligibly small,
$L_0 \approx 0$. It simplifies grain growth law to $L^m = k(t-t_0)$. Now we combine
grain growth law, Eq.~\ref{eq:grain_frac} and Eq.~\ref{eq:diff_coef_appr} to
obtain expression for the time-dependent interdiffusion coefficient:
\begin{equation}
	D(t) = D_{GB}\left[\frac{k}{(q\delta)^m}(t - t_0)\right]^{-1/m}
	\label{diff_time} \addtocounter{equation}{1} \tag{\theequation\,a}
\end{equation}

The meaning of $t_0$ needs additional specification. Grains in different regions
of the product layer begin to grow not at the same time. A grain within \textit{PL}
starts to grow not when the experiment begins, but when the grain nucleates at
the reactive boundary, that is, at the passage time, $t^*(x)$, of the reactive
boundary through the coordinate $x$. Eq.~\ref{diff_time} can now be rewritten in
terms of passage time, with the interdiffusion coefficient dependent on both time
and coordinate:
\begin{equation}
	D(x, t) = \frac{q\delta D_{GB}}{k^{1/m}} \left(t - t^*(x)\right)^{-1/m}
	\label{diff_coord_time} \tag{\theequation\,b}
\end{equation}

We expect that the movement of the reactive boundary is according to the power law 
$\ell(t) = \lambda t^{1/n}$. Assuming this to be true, we can express the passage
time as $t^*(x) = (x/\lambda)^{n}$, and the diffusion equation Eq.~\ref{eq:fick_eq_d_var}
becomes:
\begin{equation}
	\pd{C}{t} = \pd{C}{x} \frac{q\delta D_{GB}}{k^{1/m}} \frac{n}{m} \frac{1}{x}
	\left(\frac{x}{\lambda}\right)^{n}
	\left(t - \left(\frac{x}{\lambda}\right)^{n}\right)^{-\frac{1+m}{m}}   
	+ \pdd{C}{x} \frac{q\delta D_{GB}}{k^{1/m}}\left(t - \left(\frac{x}{\lambda}\right)^{n}\right)^{-1/m} 
	\label{eq:diff_eq_simpl}
\end{equation}

We search for a self-similar solution of Eq.~\ref{eq:diff_eq_simpl}, a solution
that can be expressed as a function of a single variable, given by the similarity
transformation:
\[
z = \frac{x}{\ell(t)} = \frac{x}{\lambda t^{1/n}}
\]
This solution exists, if the partial differential equation 
Eq.~\ref{eq:diff_eq_simpl} transforms to an ordinary differential equation by the
similarity transformation. Details of this transformation are given in \ref{ap:compl}.
The final result is that a condition on exponents of power-law kinetics
($n$) and grain growth ($m$) is imposed:
\begin{equation}
	\frac{1}{m} + \frac{2}{n} = 1
	\label{eq:TheLink}
\end{equation}

This equation gives a convenient means of verification of the proposed model,
as both exponents are easy to measure experimentally. Let us also examine the
limiting cases of Eq.~\ref{eq:TheLink}:
\begin{itemize}
	\item if $m=2$, then $n=4$. The value $m=2$ is the smallest	experimentally
		observed of grain growth exponent, and it is predicted by the grain-growth
		models and numerical simulations \cite{Hu1970,Louat1974,Anderson1989,
		Gao1996,Liu2001,Yu2003,Najafkhani2021}. At the same time,
		$n=4$ is the largest growth exponent observed experimentally.
	\item if $m\to \infty$, then $n\to 2$. It means that if no grain growth
		occurs, Eq.~\ref{eq:TheLink} reduces the kinetics to a diffusion control
		regime with constant interdiffusion coefficient ($\ell\propto\sqrt{t}$),
		as should be expected.
\end{itemize}

Eq.~\ref{eq:TheLink} connects exponents $n$ and $m$ and shows competition between
processes of mass transport and grain boundary movement. If grain growth is slow,
than interdiffusion coefficient is almost constant and kinetics of a reaction 
is parabolic. If grain growth is fast, the grain boundaries (routes of rapid
diffusion) move and disappear faster, so the overall reaction kinetics 
is slower.

Having made the similarity transformation, we can obtain the analytical solution
to Eqs.~\ref{eq:fick_eq_d_var}--\ref{eq:diff_bound_cond}:
\begin{numcases}{}
	 \ell(t) = \gamma\sqrt{\frac{q\delta D_{GB}}{k^{1/m}}}\, t^{1/n} \label{eq:nonpar_gr}\\
	C(x, t) = C_{max} - (C_{max} - C_{eq}) \frac{\int_{0}^{x/\lambda t^{1/n}}
		{(1 - y^n)^{1 - 2/n} \exp\left[-\frac{\gamma^2}{2n}y^2\,
F\left(\frac{2 - n}{n}, \frac{2}{n}; \frac{2+n}{n}; y^n\right)\right]\,\mathrm{d}y}}
{\int_{0}^{1} {(1 - y^n)^{1 - 2/n} 
	\exp\left[-\frac{\gamma^2}{2n}y^2\,
F\left(\frac{2- n}{n}, \frac{2}{n}; \frac{2+n}{n}; y^n\right)\right]\,\mathrm{d}y}}
\label{eq:c_anom_diff}\\
  \begin{split}
	  n(C_{max} - C_{eq})\tilde{V}_A =& \gamma^2 \exp\left[-\frac{\gamma^2}{2n}
    \,F\left(\frac{2- n}{n}, \frac{2}{n}; \frac{2+n}{n}; 1\right)\right] \\
	&\times \bigintssss_{0}^{1} {(1 - y^n)^{1 - 2/n}\exp\left[-\frac{\gamma^2}{2n}y^2\,
    F\left(\frac{2 - n}{n}, \frac{2}{n}; \frac{2+n}{n}; y^n\right)\right]
\,\mathrm{d}y} \label{eq:gamma_cond_nonpar}
  \end{split}
\end{numcases}
where $F(\frac{2 - n}{n}, \frac{2}{n};\frac{2+n}{n}; y^n)$ is a Gauss
hypergeometric series \cite{AbramowitzStegun}, given by:
\begin{equation*}
F\left(\frac{2 - n}{n}, \frac{2}{n}; \frac{2+n}{n}; y^n\right) = \frac{\Gamma(\frac{2+n}{n})}
{\Gamma(\frac{2-n}{n})\cdot\Gamma(\frac{2}{n})} \sum_{j=0}^{\infty} 
{\frac{\Gamma(\frac{2-n}{n}+j)\cdot\Gamma(\frac{2}{n} + j)}{\Gamma(\frac{2+n}{n}+j)}
\frac{y^{j\cdot n}}{j!}}
\end{equation*}

General structure of the solution is the same as in ordinary diffusion case 
(Eqs.~\ref{eq:par_gr}--\ref{eq:gamma_cond_par}), although details are more 
complicated. Eq.~\ref{eq:nonpar_gr} is equation for a width of \textit{PL} as a 
function of time; Eq.~\ref{eq:c_anom_diff} is equation for concentration within
\textit{PL} as a function of time and coordinate. Eq.~\ref{eq:gamma_cond_nonpar}
determines the numerical factor $\gamma$ that appears in Eqs.~\ref{eq:nonpar_gr}
and \ref{eq:c_anom_diff}.

\begin{figure}[h]
	 \centering
	 \includegraphics[width=0.8\linewidth]{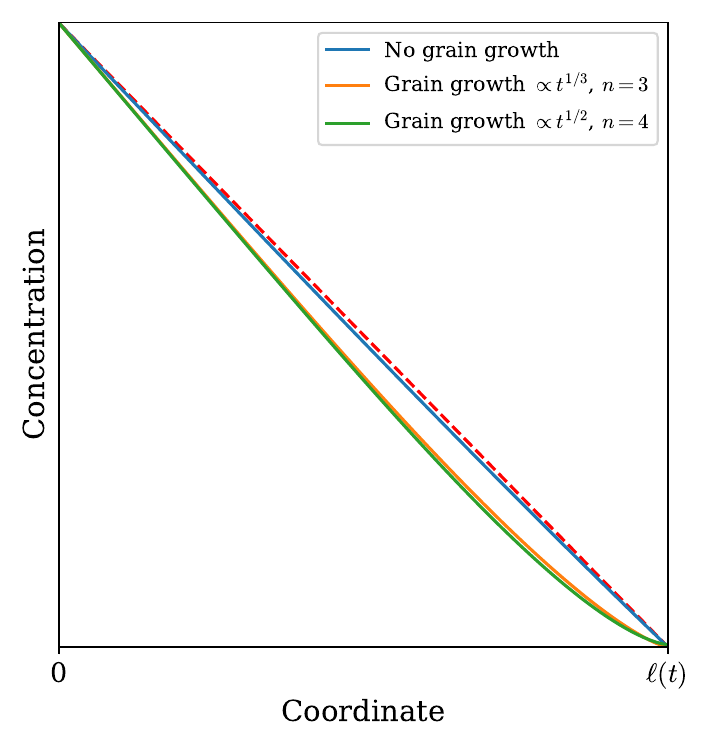}
	 \caption{General shape of concentration profiles for different kinetic of 
	 product layer growth. Straight line (red dashed) is added as a visual aid.}
	 \label{fig:prof_compare}
 \end{figure}

For the visual comparison, thee profiles (one of ordinary
diffusion case and two of diffusion accompanied with grain growth) are shown
in Fig.~\ref{fig:prof_compare}. The profiles are constructed for the same 
width of product layer. One can see how slightly ordinary diffusion profile 
differs from a straight line (red dashed line), and how profiles in cases
of nonparabolic growth are qualitatively different from it. The latter two are
also almost linear within majority of the product layer bulk, but the profiles
bend so that the slope becomes vanishingly small near the moving boundary. This
feature arises from the grains becoming ever finer near the boundary, and if
such profile is observed it should be regarded as another evidence of grain growth.

\section{Experimental}

Iridium was cut from as-received plate (99.97\% purity, GOST 55084-2012,
Russia) to size 5~\texttimes~5~\texttimes~0.1~mm for use in reaction couples.
ZrC was preliminary sintered into pellets sized \diameter 
12~\texttimes~3 mm from the powder (99.8\% purity, $\mathrm{D_{50}}$ = 2~\textmu m, technical
specifications TU 6-09-408-75) by hot pressing at 1900\textdegree C under
20~MPa for 30 min. Pellets were polished from one face using a set of diamond
suspension (Aquapol-P, Kemet Internation Limited, UK) with particle sizes 
9--1\textmu m. The density of the pellets was 6.512\textpm0.003~g~cm$^{-3}$
(measured with Upyc 1200e V5.07), 97\% of the theoretical; the lattice parameter
was 4.693~\AA (refined by Rietveld method using Topas 4.2; XRD data collected on 
D8 Advance powder diffractometer, Bruker AXS, USA), corresponding to a
stoichiometric composition. 

Reaction couples were assembled with iridium plate on top of ZrC pellet, heated 
in UGP Laboratory hot press (IA\&E SB RAS, Russia) to a given temperature (1500,
1550 or 1600\textdegree C) and held at that temperature for a holding time 
ranging from 2 to 16~h; a pressure of 5~MPa was appllied to the couple to ensure
better contact. After annealing, reaction couples were cross-sectioned, packed
with an epoxy, and polished with diamond suspension. Morphology was characterized
by scanning electron microscopy (SEM) using TM-1000 (Hitachi Ltd, Japan) and 
MIRA 3 LMU (TESCAN, Czech Republic). Measurements of product layer thickness,
carbon particles average area and average grain size were performed on aquired
microphotographs using ImageJ \cite{Schneider2012}.
Elemental composition was measured by wavelength-dispersive spectroscopy
(WDS) using JXA–8230 electron probe microanalyzer (Jeol Ltd, Japan) at an
accelerating voltage of 20~kV; PET (Zr) and LiF (Ir) were used as analyzer
crystals, pure iridium (Ir) and zirconium carbide (ZrC) were used as standards
to calculate concentrations \cite{Nikiforov2024}. Thus were obtained concentration 
profiles along reaction couples and elemental compositions in vicinity of 
boundaries Ir/\ce{ZrIr_3} and \ce{ZrIr_3}/ZrC (probes taken in 5\textmu m from
the boundary). The structure of the carbon formed within product layer was
characterized by Raman spectroscopy using a microscope LabRam HR Evolution 
(Horiba, Japan).
	
 \section{Results and Discussion}
 \subsection{Morphology and elemental composition}
\begin{figure}
	 \centering
	 \includegraphics[width=\linewidth]{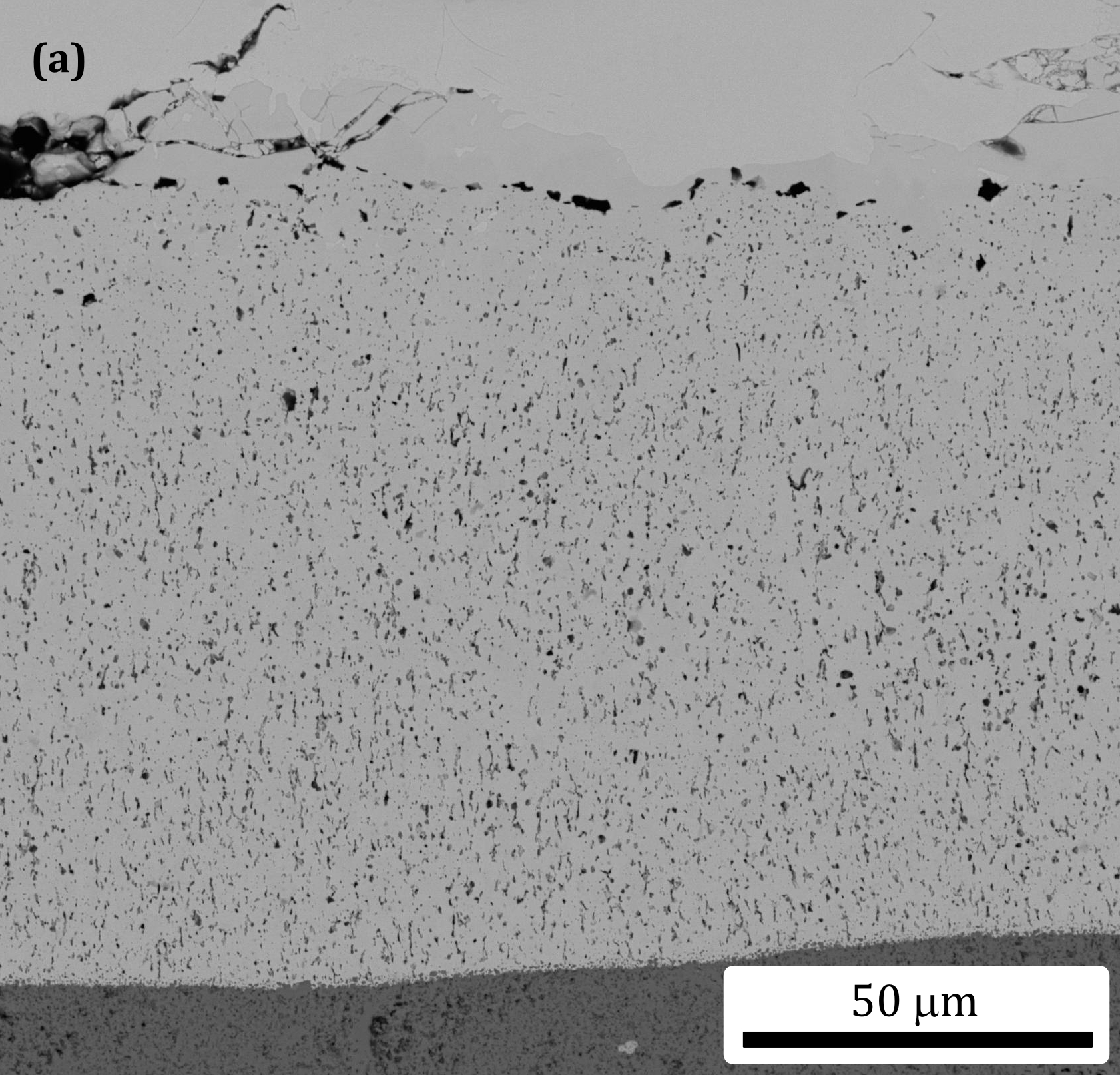} \\
	 \includegraphics[width=0.49\linewidth]{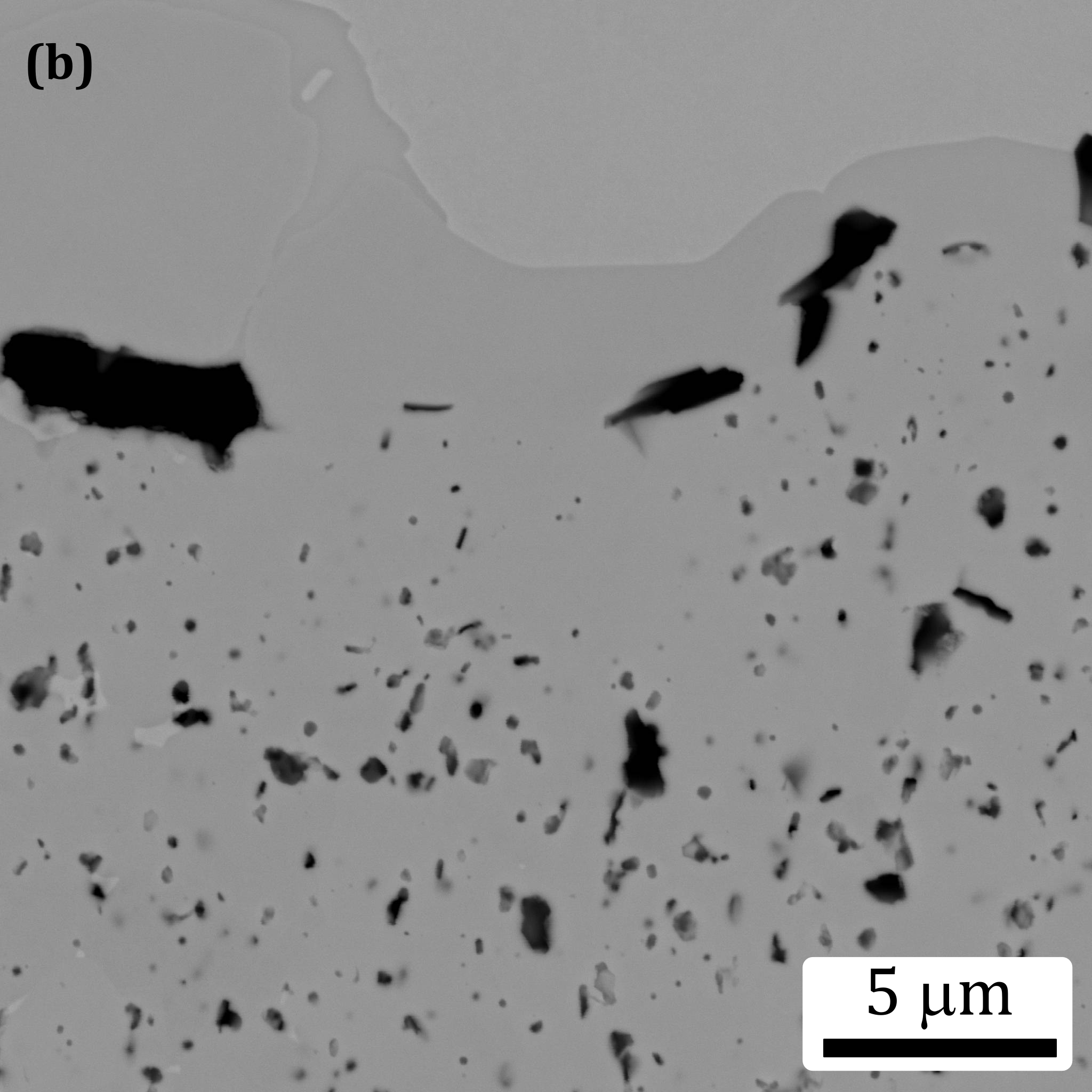}
	 \includegraphics[width=0.49\linewidth]{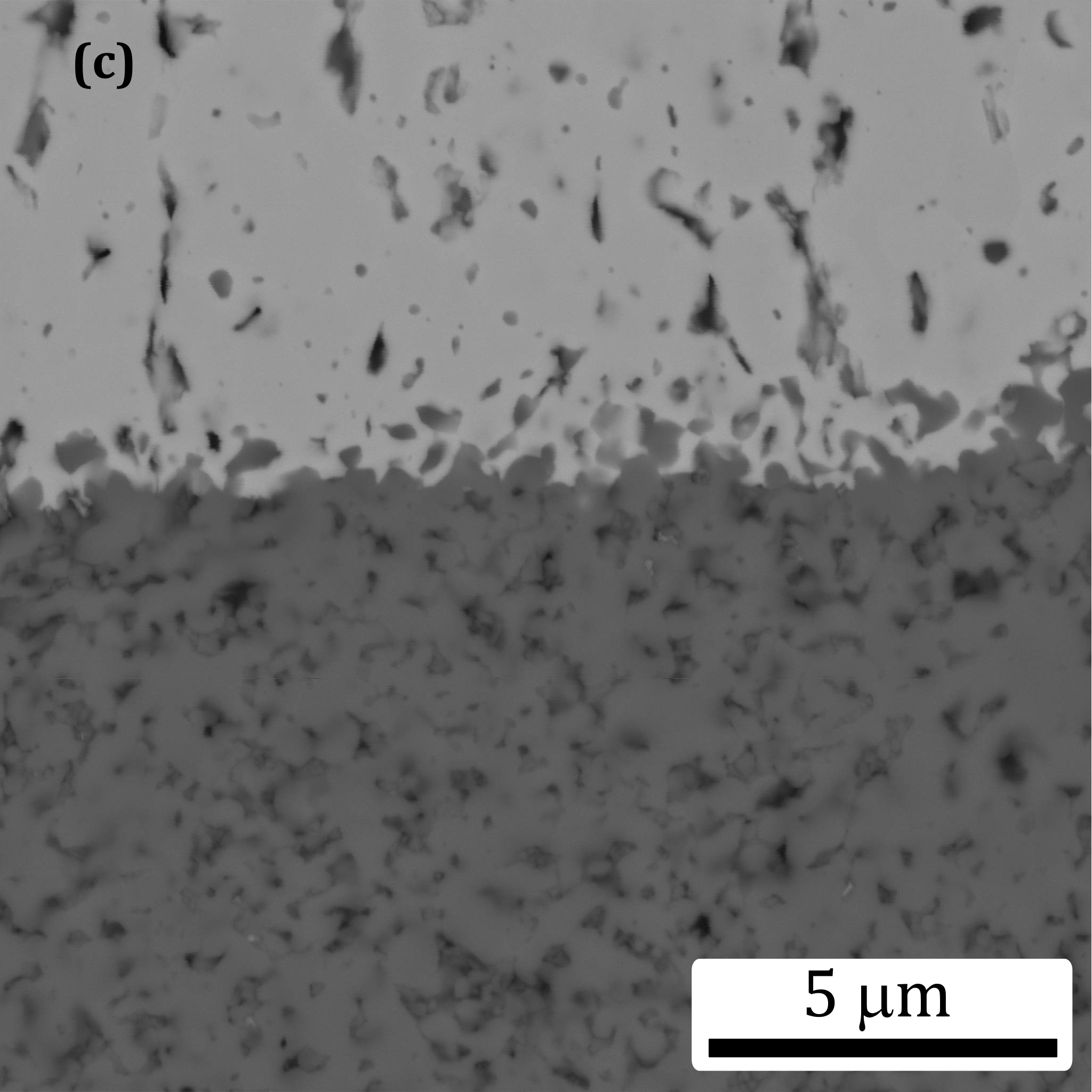}
	 \caption{Morphology of a reaction couple; sample shown is heat-treated
	 at 1600\textdegree C for 8 h. \textit{a} --- general view with all layers
 shown. \textit{b} --- image of the Ir/\ce{ZrIr_3} interface. \textit{c} --- image
of the \ce{ZrIr_3}/ZrC interface.}
	 \label{fig:dc_overview}
\end{figure}

All samples have similar features of the morphology of the 
 product layer (PL). Fig.~\ref{fig:dc_overview} shows an SEM image of a 
 cross-section of the reaction couple annealed for 8 hours at 1600\textdegree{C}.
 \textit{PL} consists of a continuous matrix of \ce{ZrIr_3} with small carbon
 inclusions. The cross-section of \textit{PL} has a compact
 polished  surface with no voids (SE-SEM images given in Supplement 1), as was
 expected from that the volume of products (per 1 mol of \ce{ZrIr_3}) exceeds 
 by 4\% the volume of reactants. Carbon inclusions are present throughout the
 \textit{PL}, with only a thin region near the Ir/\ce{ZrIr_3} interface appearing
 carbon-free.

\begin{figure}
	 \centering
	 \includegraphics[width=\linewidth]{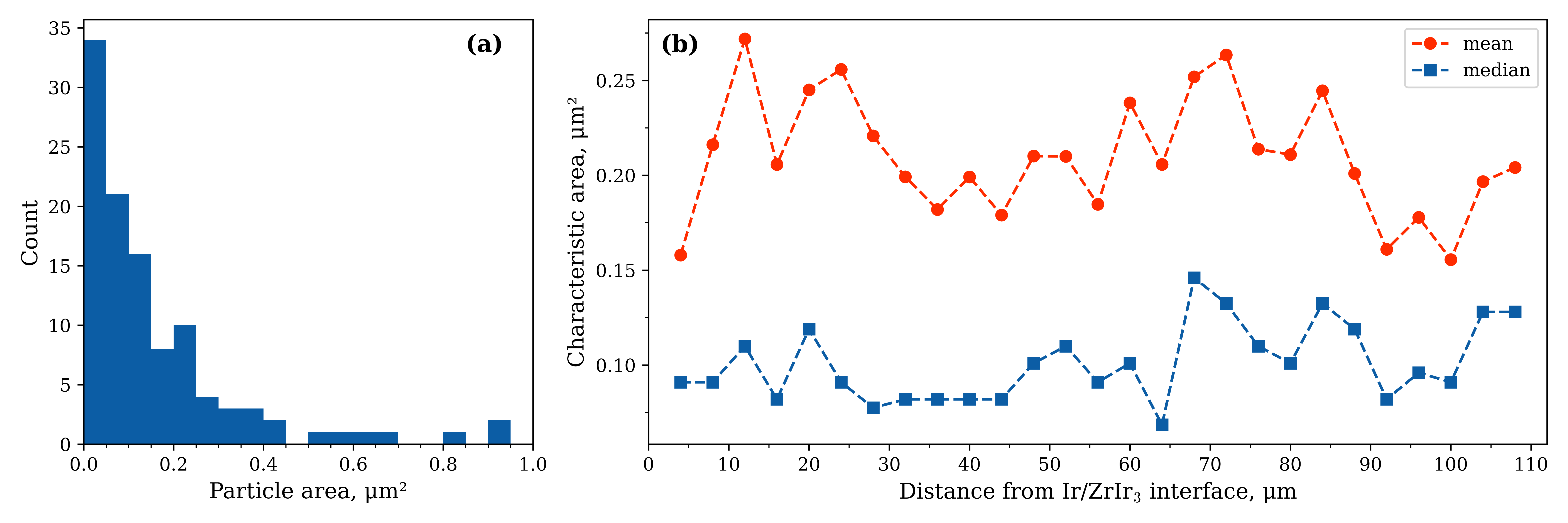} \\
	 \caption{Characterization of carbon within the product layer: \textit{a} --- size
	 distribution of carbon particles; \textit{b} --- mean and median areas of carbon particles.}
	 \label{fig:carbon_distr}
 \end{figure}

 Actual presence and structure of carbon was confirmed by Raman spectroscopy.
 The spectra from black spots throughout the \textit{PL} always showed the presence
 of peaks at \textit{ca.} 1330, 1580, and 2660~cm$^{-1}$ corresponding to D, G,
 and 2D graphite bands (see Supplement 2). To check the possibility of
 recrystallization of carbon within the \textit{PL}, the area of carbon particles
 was measured along \textit{PL} from the Ir/\ce{ZrIr_3} to \ce{ZrIr_3}/ZrC interface.
 Cross-sectional areas of carbon particles are distributed 
 exponentially, with the number of particles decaying with an increase in size
 (Fig.~\ref{fig:carbon_distr}a). Such distribution is probably caused by the
 process of formation of carbon particles on the reactive boundary, with small
 particles having a greater probability of formation than the larger ones. Once
 formed, the carbon particle would not change its size, as can
 be inferred from Fig.~\ref{fig:carbon_distr}b, showing mean and median 
cross-section areas of carbon particles.

Thus, carbon can be regarded as an inert indicator of the
reaction at the \ce{ZrIr_3}/\ce{ZrC} interface. As noted earlier,
the carbon is observed throughout the \textit{PL}, which means the reaction proceeds
via reaction with at the \ce{ZrIr_3}/\ce{ZrC} interface, while Ir/\ce{ZrIr_3}
boundary acts as only a source of iridium for the intermetallic phase.

\begin{figure}
	 \centering
	 \includegraphics[width=0.75\linewidth]{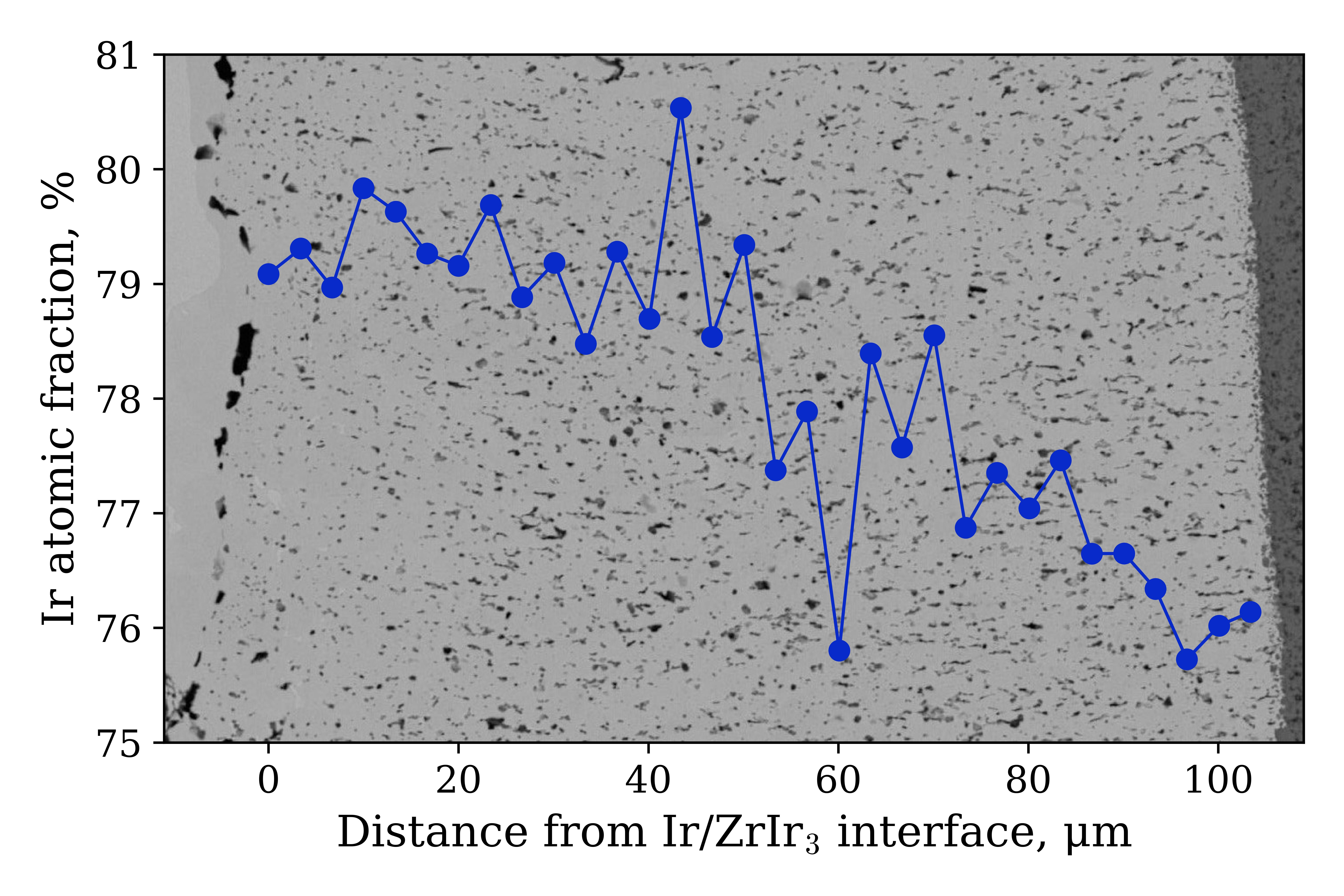} \\
	 \caption{Concentration profile of iridium across the \ce{ZrIr_3} phase.}
	 \label{fig:conc_prof}
 \end{figure}

The intermetallic \ce{ZrIr_3} phase adopts L$_1$2 structure type and can form a
range of substitutional solid solutions \cite{Ran2006}, both sub- and
over-stoichiometric by iridium. Fig.~\ref{fig:conc_prof} shows a concentration
profile in a region of a reaction couple heat-treated for 8 hours at 
1600\textdegree C; other samples show similar features. One can see, that indeed
the \textit{PL} contains only one intermetallic compound --- \ce{ZrIr_3}. This
phase is overstoichiometric by Ir, in accordance with previous investigations
with powder samples \cite{Nikiforov2024, Brewer1973}. Composition near the
 boundary Ir/\ce{ZrIr_3} is close to the upper limit of the homogeneity range
  of \ce{ZrIr_{3\textpm x}}, while the composition near the 
boundary \ce{ZrIr_3}/ZrC does not reach values of Ir atomic fraction less than 75\%.
The seeming  absence of the substoichiometric \ce{ZrIr_3} suggests that the
equilibrium composition of \ce{ZrIr_3} in contact with ZrC lies close to 75~mol.\%.

\subsection{Kinetics of the product layer growth}

\begin{figure}
	 \centering
	 \includegraphics[width=0.65\linewidth]{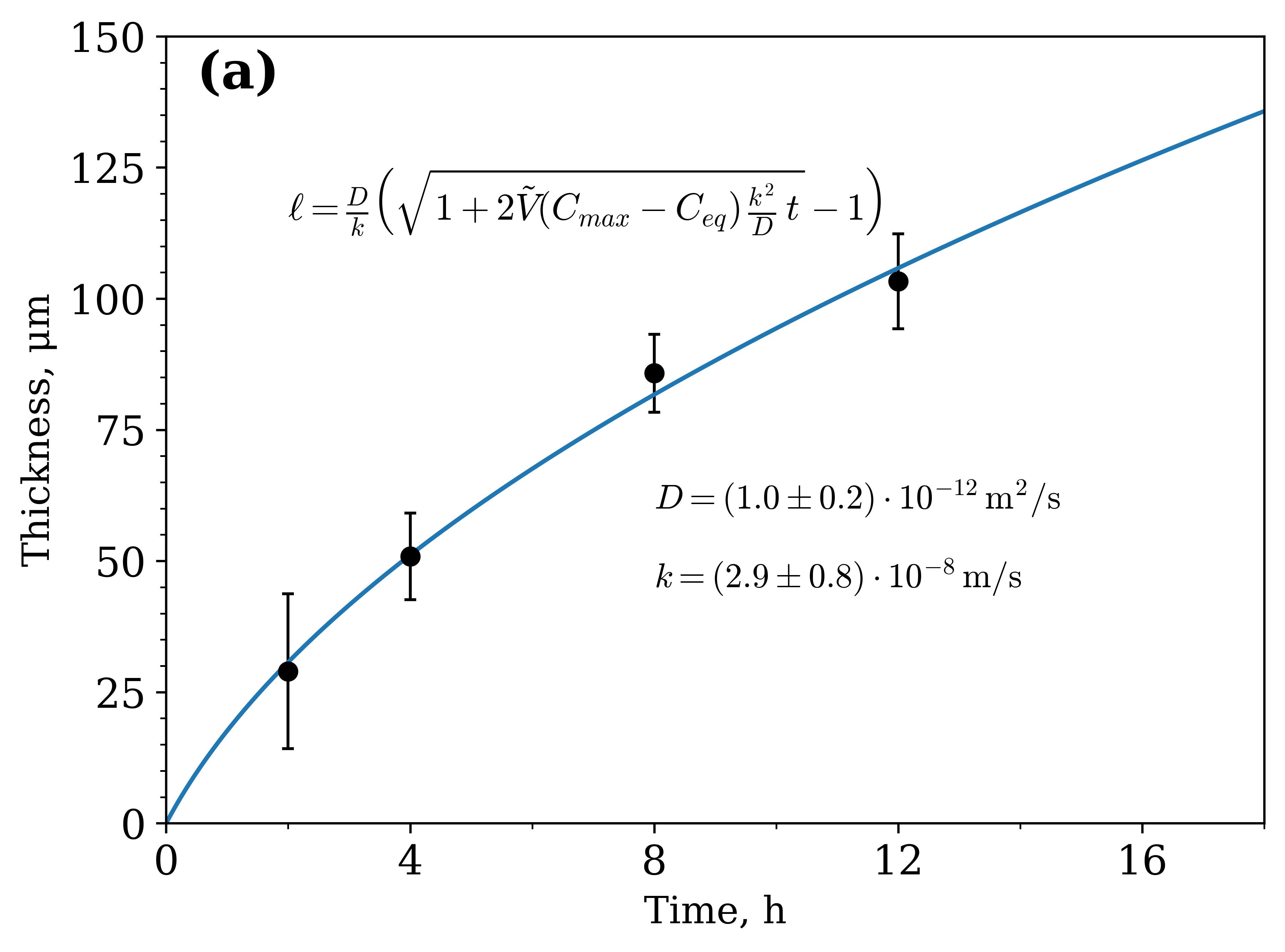}
	 \includegraphics[width=0.65\linewidth]{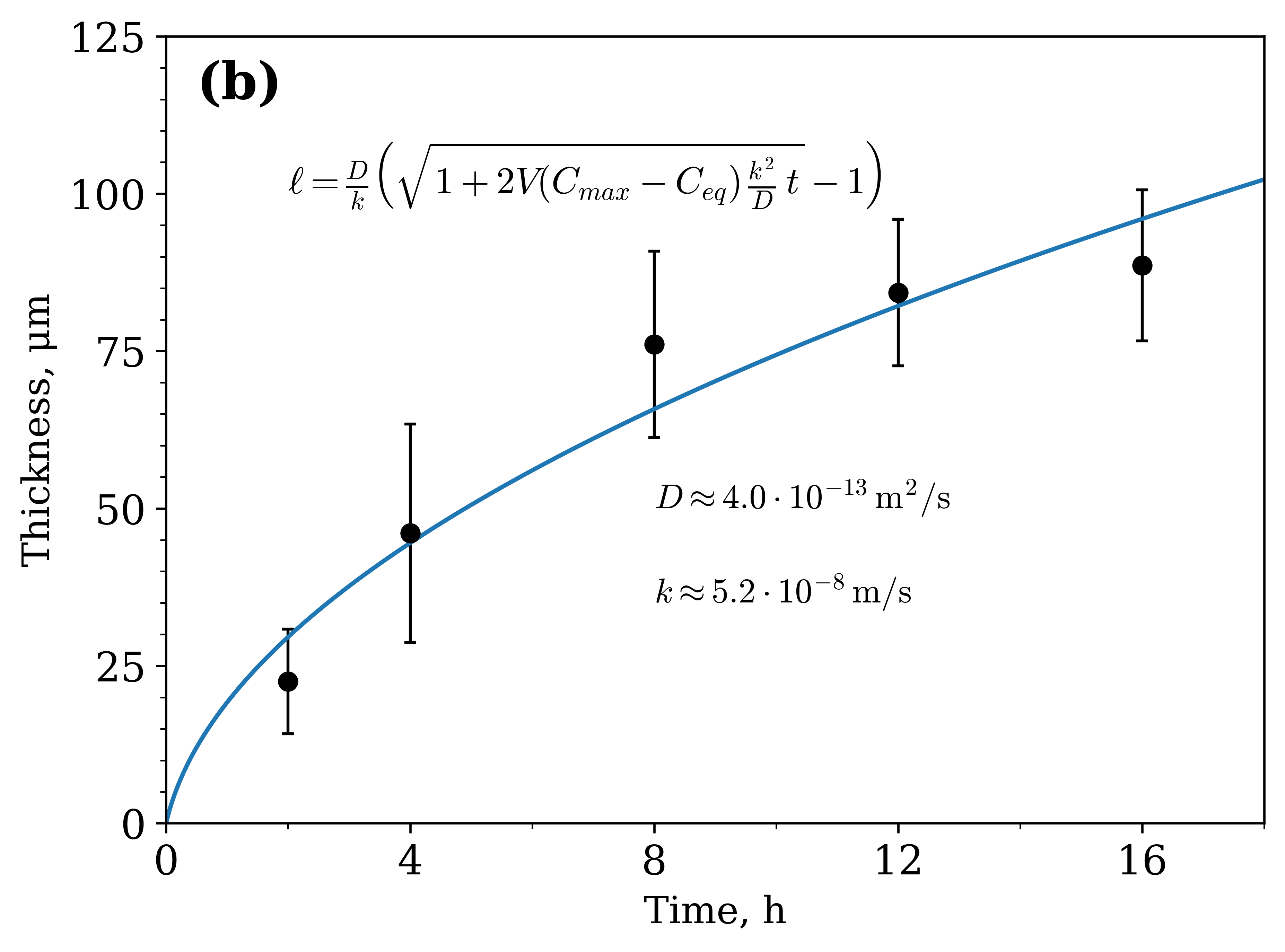}
	 \includegraphics[width=0.65\linewidth]{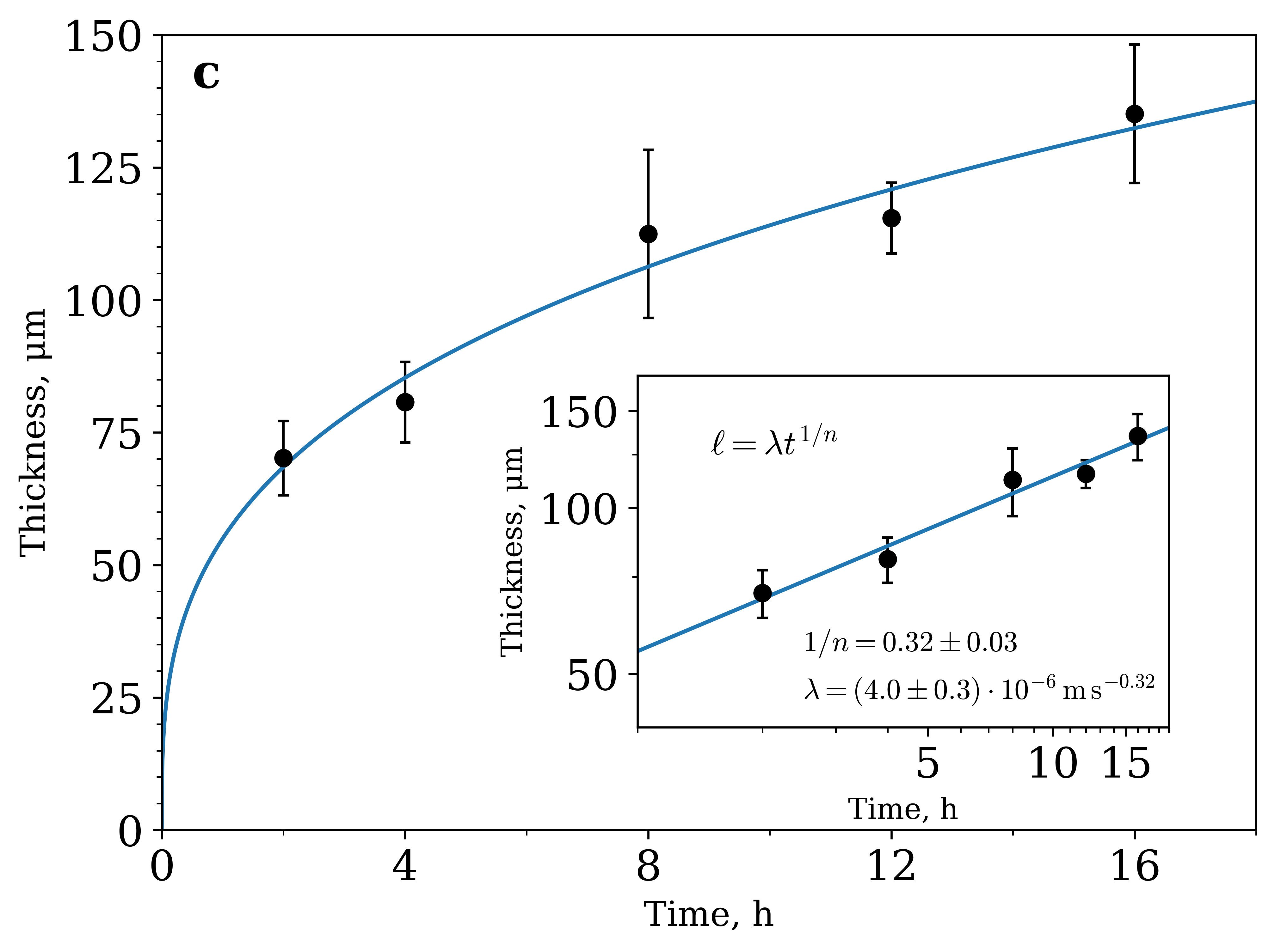}
	 \caption{Thickness of the \ce{ZrIr_3}--C layer versus time at three
	 different temperatures: \textit{a} --- 1500\textdegree C; \textit{b} --- 
     1550\textdegree C; \textit{c} --- 1600\textdegree C.}
	 \label{fig:all_temps_growth}
 \end{figure}

The plot of the mean thickness of the \textit{PL} versus time is shown in
Fig.~\ref{fig:all_temps_growth}, with error bars showing the standard deviation
of the thickness of the \textit{PL}. One can see that the thickening behavior
gradually changes with an increase in temperature from 1500 to 1600\textdegree C. 

First, we consider samples treated at 1500 and 1550\textdegree C because they show
similar growth kinetics (Fig.~\ref{fig:all_temps_growth}a, b). Fitting experimental
points with existing models showed that the kinetics at these temperatures is 
best described by Eq.~\ref{eq:react_control}, so the the reaction at the
\ce{ZrIr_3}/ZrC interface controls the growth of the \textit{PL}. 

Fitting the \textit{PL} thickness plots by Eq.~\ref{eq:react_control} gives 
the following values of the interdiffusion coefficient and reaction rate coefficient:
$D = (1.0\pm 0.2)\cdot 10^{-12}\text{ m}^2/\text{s}$
and $k = (2.9\pm 0.7)\cdot10^{-8}\text{m/s}$ (at 1500\textdegree C); 
$D = (4.0\pm 1.5)\cdot 10^{-13}\text{ m}^2/\text{s}$ and 
$k = (5\pm8)\cdot10^{-8}\text{m/s}$ (at 1550\textdegree C).
The points at 1500\textdegree C are satisfactorily approximated,
while the points at 1550\textdegree C are approximated poorly, as can be seen from 
Fig.~\ref{fig:all_temps_growth} and large uncertainties in values of $D$ and $k$.
Thus, at 1500\textdegree C, kinetics is controlled by the interface reaction, 
while at 1550\textdegree C, the reaction is in a transition regime that is not
described well by any model considered in this work.

It is worth noting that the values of the interdiffusion
coefficient indicate that diffusion of Ir atoms occurs along grain boundaries.
It was shown earlier \cite{Ikeda1998,Numakura2001,Uchida2005} that the bulk
interdiffusion coefficient in L$_1$2 alloys is by 1--2 orders of magnitude less
than the self-diffusion coefficient in a parent metal at the same reduced temperature
($T/T_m$). At 1616\textdegree C (corresponding to reduced temperature $T/T_m = 0.694$
of \ce{ZrIr_3} at 1500\textdegree C) and 1669\textdegree C (corresponding to 
$T/T_m = 0.714$ of \ce{ZrIr_3} at 1550\textdegree C) self-diffusion coefficient
of iridium equals $2.64\cdot10^{-17}$ and $5.67\cdot10^{-17}\text{ m}^2/\text{s}$
accordingly \cite{Ran2006,Mehrer1990}. The interdiffusion coefficients obtained
in this work are five orders of magnitude greater than the self-diffusion
coefficient of pure iridium, not less than it. That is characteristic of 
the grain-boundary diffusion, which typically exceeds bulk diffusion by 4--6
orders of magnitude \cite{Mishin1997}. 

A similar feature can be noticed in the earlier work of Kwon \cite{Kwon1989}. 
This work investigated a reaction between hafnium carbide and iridium at 
temperatures 1900--2200\textdegree C. Kwon found that a product layer
(\ce{HfIr_3} + C) grew parabolically and calculated the tracer diffusion
coefficient of Ir in \ce{HfIr_3} (which contributes the most to the interdiffusion
coefficient) from the kinetic data. For instance, it was found that 
$D_\text{Ir}^* = 2.74\cdot10^{-14}$ at 2000\textdegree C, while the self-diffusion
coefficient in elemental Ir at 1981\textdegree C (same $T/T_m = 0.828$) is 
$2.44\cdot10^{-15}$. Again, the diffusion in the intermetallic compound appears
to be faster than bulk diffusion in pure metal, indicating the grain-boundary
character of diffusion. Such behavior might likely be typical for such systems,
where the products (intermetallic compound and carbon) form one layer. However,
the degree of grain-boundary influence should depend on a particular system.

Secondly, we consider samples annealed at 1600\textdegree C. The kinetics of the
product layer growth at 1600\textdegree C differs from kinetics at 1500 and 1550\textdegree C 
(Fig.~\ref{fig:all_temps_growth}c) and is described by the power law. Fitting
experimental points with Eq.~\ref{eq:nonpar_gr} gives values of exponent 
$n = 0.32\pm0.03$ and growth constant 
$\lambda = (4.0\pm0.3)\cdot10^{-6}\text{ m/s}^{0.32}$.

\begin{figure}
	 \centering
	 \includegraphics[width=\linewidth]{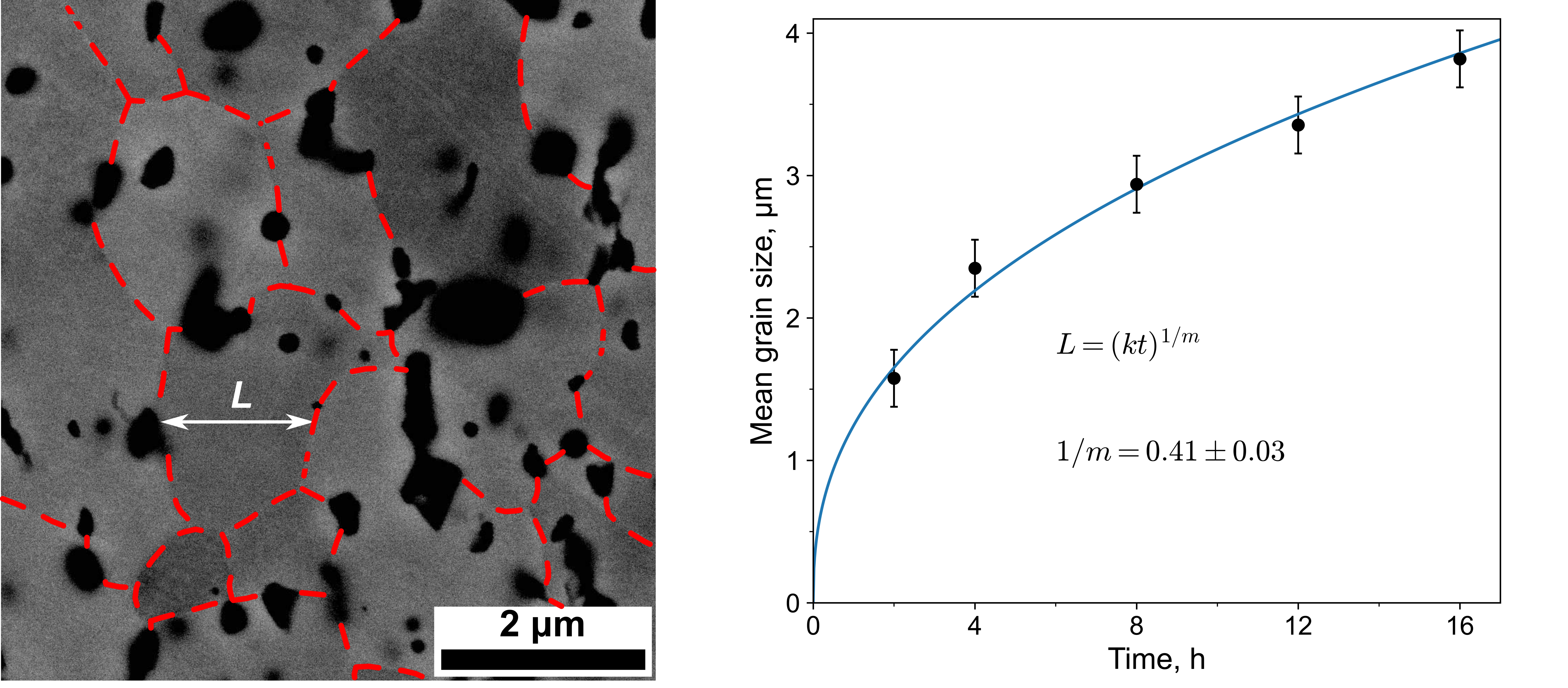}
	 \caption{Average grain size in the vicinity of Ir/\ce{ZrIr_3}
	 boundary: \textit{a} --- delineation of grain boundaries and measurement of 
 cross-sectional lenth; \textit{b} --- plot of average grain size vs. time.}
	 \label{fig:grain_growth}
 \end{figure}

In Section~\ref{ssec:gg_kinetics}, we have argued that
non-parabolic growth should be the consequence of grain growth in the \textit{PL},
where diffusion occurs. In the case of the reaction Ir+ZrC, the grain growth
should be observable in the \ce{ZrIr_3} phase. We have measured the average grain
size of \ce{ZrIr_3} to support this hypothesis. Because the \textit{PL} grows
during the experiment, it is crucial to find a region where the grain size can
be correlated unambiguously with the time. That is why the measurements were
taken in the vicinity of the Ir/\ce{ZrIr_3} interface. This region forms in the
initial stages of the reaction and the time of its evolution only slightly
differs from the time of the experiment. To find the average size, we used
highly contrasted SEM images, then delineated the grain boundaries, measured
cross-sectional lengths of grains (Fig.~\ref{fig:grain_growth}), and took the
mean value. As the plot in Fig.~\ref{fig:grain_growth}b shows, grain growth
obeys the power law Eq.~\ref{eq:grain_growth} with initial grain size set to
zero, following the approximation adopted in Section~\ref{ssec:gg_kinetics}.
The growth exponent was found to be $m = 0.41\pm0.03$. Furthermore, if the model
from \ref{ssec:gg_kinetics} is correct, than the Eq. \ref{eq:TheLink} should
hold. Indeed:
\begin{equation*}
\frac{1}{m} + \frac{2}{n} = (0.41\pm0.03) + 2\cdot(0.32\pm0.03) = 1.05\pm0.07
\end{equation*}
the left-hand sum in Eq.~\ref{eq:TheLink} calculated from the experimentally 
measured values of $n$ and $m$ differs from the theoretical value within
the uncertainty.

 The kinetics of the reaction \ce{ZrC + 3Ir \rightarrow ZrIr_3 + C} at studied 
temperatures can be summarized as follows. The interaction proceeds via reaction 
at the interface \ce{ZrIr_3}/ZrC, with iridium atoms diffusing through the 
\textit{PL} from the Ir/\ce{ZrIr_3} boundary. The local equilibrium is reached
at the latter boundary, and the products do not form here. At $T<$1500\textdegree C,
and supposedly lower temperatures too, fine grains of \ce{ZrIr_3} form. 
Boundaries of these fine grains are paths of rapid diffusion, so that the
overall kinetics is controlled by the reaction at the interface \ce{ZrIr_3}/\ce{ZrC}.
As temperature rises, the grains start to grow, decreasing the number of grain
boundaries and slowing down the diffusion. At 1550\textdegree C diffusion is
still fast enough, so the kinetics can be regarded as interfacial reaction 
controlled, but the effect of slowdown becomes noticeable. At 1600\textdegree C,
the transition to the diffusion-controlled regime occurs, and because the grain
growth of \ce{ZrIr_3} occurs as well, the \textit{PL} kinetics obeys the power
law with the \textit{PL} thickness proportional to $t^{0.32}$. 

Interestingly, grain growth starts after the mobility of atoms has become 
noticeably high. Grain growth is commonly associated with the mobility of atoms, but in
this case, these phenomena are distinct. It seems plausible that carbon 
inclusions play some role in it. Carbon inclusions, although not involved in
the mass transport themselves, interact with grain boundaries and act as pinning
particles \cite{Najafkhani2021}, slowing down the movement of grain boundaries.
Moreover, the grain-growth exponent $m = 1/0.41 \approx 2.4$ is greater than
"ideal" $m=2$, which is highly suggestive of the presence of the pinning effect
\cite{Grey1973}.

We think it would be profitable to study further the effects of grain growth on
reaction kinetics in order to fully understand solid-state reactions similar to
the one discussed in this work. Here, it turned out that assumptions made in
developing the theoretical model were accurate enough to describe experimental
data, but one cannot expect it always to be true. It is worth exploring how
accounting for non-zero initial grain size, bulk diffusion, and even 
time-dependent grain growth exponent (which could be observable in some cases
\cite{Liu2001,Kim2009}) would change the kinetics of a solid-state reaction.

\section{Conclusion}

In this work, we have studied the solid state reaction between iridium and 
zirconium carbide at temperatures 1500--1600\textdegree C. During the reaction,
a product layer consisting of \ce{ZrIr_3} and graphite-like carbon forms. Growth
of the product layer occurs via the reaction at the \ce{ZrIr_3}/ZrC interface,
while Ir atoms diffuse from the Ir/\ce{ZrIr_3} interface through the product
layer towards the reaction zone. The carbon acts as an inert
secondary phase that forms at the reactive boundary and does not participate
in mass transport afterward.

At 1500 and 1550\textdegree C, the interaction is in the regime of interfacial
reaction control, while at 1600\textdegree C, the kinetics is non-parabolic and
cannot be properly described as limited by the interfacial reaction or diffusion.
This work presents a theoretical treatment of the model of such non-parabolic
kinetics. This model accounts for grain growth and its effect on the explicit
time dependence of the interdiffusion coefficient. We have obtained full solution
for diffusion equation in a growing layer with time-dependent diffusion 
coefficient and derived the equation that relates grain-growth exponent and
the power-law exponent of non-parabolic kinetics. We have shown that the experimental
values of these exponents confirm that equation.

Grain boundaries play a significant role in the studied reaction regardless
of the kinetics regime. At lower temperatures, boundaries of fine grains of
\ce{ZrIr_3} provide a path of fast diffusion for iridium atoms, which manifests
in the large value of interdiffusion coefficient. 

Analysis of the reaction \ce{ZrC + 3Ir -> ZrIr_3 + C} could be useful beyond
the description of the specific system. It provides insight into the behavior
of similar systems and what one could expect of them. The role
of carbon is essential in such reactions. Although it is an inert secondary
phase, it pins grain boundaries of the intermetallic compound, thus slowing
down grain growth and ultimately leading to a qualitatively different kinetic
regime. Moreover, the theoretical framework presented here will help to 
understand and further improve our knowledge of interconnection between
microstructure and solid-state processes.

\section*{Acknowledgements}

This study was supported by the Russian Science Foundation (project no. 23-19-00212).

	\appendix
	\section{Derivations of results in Theory section}
	\label{ap:sols}
	\subsection{Interface reaction control}

	For the regime of kinetics when the reaction rate is controlled by the 
	interface reaction, the set of the equations \ref{eq:fick_eq_d_const}--\ref{eq:width_ic}
	becomes:

\begin{gather}
	 \pdd{C}{x} = 0, \quad 0\leq x\leq \ell(t)\\
	C(0, t) = C_{max} \\
	\frac{1}{\tilde{V}_A}\od{\ell}{t} = k \cdot \left(C(\ell) - C_{eq}\right) 
 = -D \left( \pd{C}{x} \right) \bigg|_{x=\ell} \\
 \ell(0) = 0 
\end{gather}

As $\pdd{C}{x}=0$, the first derivative of concentration with respect to coordinate
is uniform throughout the \textit{IL}, hence
\begin{equation*}
	\left( \pd{C}{x} \right) \bigg|_{x=\ell} = -\frac{C_{max} - C(\ell)}{\ell}
\end{equation*}
Next, we use the equality between diffusive flux and flux due to interface reaction
in (A.3) and
\begin{equation}
	k\left(C(\ell) - C_{eq}\right) = \frac{D}{\ell}\left(C_{max} - C(\ell)\right)
	\quad \Rightarrow \quad
	C(\ell) = \frac{kC_{eq} + D/\ell \cdot C_{max}}{k + D/\ell} \label{eq:c_el_ap}
\end{equation}

Subsequent substitution into the condition on boundary motion in (A.3) gives:
\begin{equation}
\frac{1}{\tilde{V}_A}\od{\ell}{t} = k \cdot 
\left(\frac{kC_{eq} + D/\ell \cdot C_{max}}{k + D/\ell} - C_{eq}\right) = 
\frac{kD/\ell}{k + D/\ell}\left(C_{max} - C_{eq}\right)
\label{eq:kin_cont}
\end{equation}
The equation \ref{eq:kin_cont} can then be solved by separation of variables, 
application of initial condition (A.4) and taking the positive branch of 
the solution:
\begin{equation}
	 \ell(t) = \frac{D}{k} \left( \sqrt{1 + 2(C_{max} - C_{eq})\tilde{V}_A \frac{k^2}{D} t} - 1 \right) 
	 \label{eq:el_kin_ap}
\end{equation}

Solution for concentration is obtained using the fact, that $C(x)$ is linear:
\begin{equation}
	C(x) = C_{max} - \frac{C_{max} - C(\ell)}{\ell}x \label{eq:ap1}
\end{equation}
Substituting Eqs.~\ref{eq:c_el_ap} and \ref{eq:el_kin_ap} into Eq.~\ref{eq:ap1},
we get the following expression for $C(x, t)$:
\begin{equation}
	C(x, t) = C_{max} -
	\frac{(C_{max} - C_{eq})x}{\sqrt{1 + 2(C_{max} - C_{eq})\tilde{V}_A \frac{k^2}{D} t}} 
\end{equation}

\subsection{Diffusion control}

In the diffusion controlled regime the set of the equations \ref{eq:fick_eq_d_const}--\ref{eq:width_ic}
becomes:
	\begin{gather}
		\pd{C}{t} = D\pdd{C}{x}, \quad 0\leq x\leq \ell(t) \label{eq:fick_ap2}\\
		C(0, t) = C_{max},\quad C(\ell, t) = C_{eq} \label{eq:bnd_cond_ap2}\\
	\frac{1}{\tilde{V}_A}\od{\ell}{t} = -D \left( \pd{C}{x} \right) \bigg|_{x=\ell}
	\label{eq:s_bnd_cond_ap}
	\\
 \ell(0) = 0 
	\end{gather}

	This system of equations is solved via a similarity transformation:
	\begin{equation}
		z = \frac{x}{\lambda t^{1/2}} \label{eq:sim_tr_ap}
	\end{equation}
	with $\lambda$ defined such that $\ell(t) = \lambda t^{1/2}$. Then we transform 
	partial derivatives $\partial C / \partial t$ and $\partial^2 C/\partial x^2$ into ordinary
	derivatives $\mathrm{d}C/\mathrm{d}z$ and $\mathrm{d}^2C/\mathrm{d}z^2$:
	\begin{gather*}
		\pd{C}{t} = -\frac{z}{2 t}\od{C}{z}\\
		\pdd{C}{x} = \frac{z^2}{x^2}\odd{C}{z} = \frac{1}{\lambda^2 t}\odd{C}{z}
	\end{gather*}

	Then Eq. \ref{eq:fick_ap2} transforms into:
	\begin{equation}
		-\frac{\lambda^2 z}{2D}\od{C}{z} = \odd{C}{z}
	\end{equation}

	Integration of this equation with boundary conditions (Eq.~\ref{eq:bnd_cond_ap2})
	leads to the following solution:
	\begin{equation}
		C(x, t) = C_{max} - \frac{C_{max} - C_{eq}}
		{\operatorname{erf}(\lambda/2\sqrt{D})}\operatorname{erf}\left(\frac{x}{2\sqrt{Dt}}\right)
	\end{equation}

At this point it is convenient to introduce a dimensionless parameter
$\gamma=\lambda/2\sqrt{D}$, so that the boundary moves according to equation:
\begin{equation}
	\ell = 2 \gamma\sqrt{Dt}
\end{equation}
and Eq.~A.14 becomes:
	\begin{equation}
		C(x, t) = C_{max} - \frac{C_{max} - C_{eq}}
		{\operatorname{erf}(\gamma)}\operatorname{erf}\left(\frac{x}{2 \sqrt{Dt}}\right)
	\end{equation}
	Parameter $\gamma$ can be found with the use of Eq. \ref{eq:s_bnd_cond_ap}.
	After taking derivatives $\mathrm{d}\ell/\mathrm{d}t$ and 
	$\partial C/\partial x$ and substituting it into Eq. \ref{eq:s_bnd_cond_ap}
	we obtain the following equation on $\gamma$, from which it can be determined:
	\begin{equation}
		\frac{(C_{max} - C_{eq})\tilde{V}_A}{\sqrt{\pi}} = \gamma \operatorname{erf}(\gamma)
		\exp(\gamma^2)
	\end{equation}
	
\subsection{Grain-growth accompanied kinetics}
\label{ap:compl}

This appendix is designed to give a more complete treatment of the model 
discussed in Section~\ref{ssec:gg_kinetics}. First, we will derive a more general diffusion
equation when the grain growth is present. Then we will show how the two simplifying
assumptions lead to the result obtained earlier in Section~\ref{ssec:gg_kinetics}. And 
lastly, we will show, how the solution to the diffusion equation is obtained.
The notation is the same as in Section~\ref{ssec:gg_kinetics}.

The system of equations is:
\begin{gather}
	\pd{C}{t} = \pd{}{x}\left(D(x,t)\pd{C}{x}\right), \quad 0\leq x\leq \ell(t) \label{eq:fick_eq_d_var_ap}\\
	C(0, t) = C_{max}, \quad C(\ell, t) = C_{eq} \label{eq:bnd_cnd_ap}\\
	\frac{1}{\tilde{V}_A}\od{\ell}{t}  = -D \left( \pd{C}{x} \right)
	\bigg|_{x=\ell} \label{eq:stef_bc_ap}\\
 \ell(0) = 0 
\end{gather}
Time dependence of interdiffusion coefficient can be found with the use of the 
following equations:
\begin{gather}
	D = g D_{GB} +  (1-g)D_B \label{eq:diff_eff_ap} \\
	g = \frac{q\delta}{L} \label{eq:grain_ap} \\
	L^m - L_0^m = k(t - t_0) \label{eq:grain_growth_ap}
\end{gather}

Combining Eqs.~\ref{eq:grain_ap} and \ref{eq:grain_growth_ap} and solving for 
$g(t)$, we obtain:
\begin{equation}
	g(t) = \left[g_0^{-m} + \frac{k}{(q\delta)^m}(t - t_0)\right]^{-1/m}
	\label{eq:g_time_ap}
\end{equation}

Inserting it into Eq.~\ref{eq:diff_eff_ap} gives the time dependence of the 
effective interdiffusion coefficient:
\begin{equation}
	D(x,t) = D_B + (D_{GB} - D_B)\left[g_0^{-m} + \frac{k}{(q\delta)^m}(t - t^*(x))\right]^{-1/m}
	\label{eq:diff_time_ap}
\end{equation}
where $t^*(x)$ is the time when the moving boundary passes through the coordinate
$x$ (passage time); in our case this passage time and time $t_0$ (start of grain
growth) coincide.

The next step is to use the expression of the time-dependent interdiffusion
coefficient to find the explicit diffusion equation. To do that, we rewrite 
Eq.~\ref{eq:fick_eq_d_var_ap} as:
\begin{equation}
	\frac{\partial C}{\partial t} = \frac{\partial D}{\partial x}\frac{\partial C}{\partial x} + D\frac{\partial^2 C}{\partial x^2}
	\label{diff_eq}
\end{equation}
Then, we express the derivative $\frac{\partial D(x,t)}{\partial x}$:
\begin{equation}
	\frac{\partial D(x,t)}{\partial x} = \frac{k(D_{GB} - D_B)}{m(q\delta)^{m}}
	\left[g_0^{-m} + \frac{k}{(q\delta)^m}(t - t^*(x))\right]^{-\frac{1+m}{m}} \frac{dt^*(x)}{dx}
	\label{eq:diff_coeff_der_ap}
\end{equation}
The diffusion equation (\ref{diff_eq}) thus becomes:
\begin{multline}
	\frac{\partial C}{\partial t} =  \frac{k(D_{GB} - D_B)}{m(q\delta)^{m}}
	\left[g_0^{-m} + \frac{k}{(q\delta)^m}(t - t^*(x))\right]^{-\frac{1+m}{m}}
	\frac{dt^*(x)}{dx}\frac{\partial C}{\partial x}
	\\ + \left[D_B + (D_{GB} - D_B)\left(g_0^{-m} + \frac{k}{(q\delta)^m}(t - t^*(x))
	\right)^{-1/m}\right]
	\frac{\partial^2 C}{\partial x^2} 
	\label{diff_eq_big}
\end{multline}

Now we use some assumptions to simplify Eq.~\ref{diff_eq_big}. We expect that 
the movement of the boundary is governed by the power law $\ell(t) = \lambda t^{1/n}$.
So we can express the passage time as $t^*(x) = (x/\lambda)^{n}$. Inserting it
into Eqs.~\ref{eq:diff_time_ap} and \ref{eq:diff_coeff_der_ap} we can express 
the interdiffusion coefficient and its partial derivative $\partial D/\partial x$ as:
\begin{align}
&	D(x, t) = D_B + (D_{GB} - D_B)\left[g_0^{-m} + \frac{k}{(q\delta)^m}\left(t - 
		\left(\frac{x}{\lambda}\right)^{n}\right)\right]^{-1/m} \addtocounter{equation}{1}
	\label{diff_ap} \tag{\theequation\,a}
	\\
&	\pd{D(x,t)}{x} = \frac{nk(D_{GB} - D_{B})}{m(q \delta)^{m}x}
    \left(\frac{x}{\lambda}\right)^{n} \left[g_{0}^{-m}
	+ \frac{k}{(q\delta)^m}\left(t - \left(\frac{x}{\lambda}\right)^{n}\right)\right]
	^{-\frac{1+m}{m}} \label{diff_pr_ap} \tag{\theequation\,b}
\end{align}

The next assumptions are:
\begin{itemize}
	\item Grains in the vicinity of the reactive boundary are substantially 
		smaller than grains in the rest of the \textit{PL}: $L_0 \to 0$. 
		Because $g_0 \propto L_0^{-1}$, we can neglect term $g_0^{-m}$ in square
		brackets in Eqs.~\ref{diff_ap} and \ref{diff_pr_ap}.

	\item Bulk diffusion can be neglected in comparison with the grain-boundary
		diffusion: $D \approx g D_{GB}$.
\end{itemize}

Given the aforementioned assumptions, we simplify (\ref{diff_eq_big}) to:
\begin{multline}
	\pd{C}{t} =   \frac{n k D_{GB}}{m(q \delta)^{m}x}\left(\frac{x}{\lambda}\right)^{n}
	\left[\frac{k}{(q\delta)^m}\left(t - \left(\frac{x}{\lambda}\right)^{n}\right)
	\right]^{-\frac{1+m}{m}}\pd{C}{x}  \\ 
	+  D_{GB} \left[\frac{k}{(q\delta)^m}\left(t - \left(\frac{x}{\lambda}\right)^{n}
		\right)\right]^{-\frac{1}{m}}\pdd{C}{x} 
	\label{eq:diff_eq_simpl_ap}
\end{multline}

Eq.~\ref{eq:diff_eq_simpl_ap} can be solved in a way similar to the solution of
Eq.~\ref{eq:fick_ap2} --- introducing a similarity transformation:
\[
z = \frac{x}{\lambda t^{1/n}}
\]	
Partial derivatives, then, are transformed as follows:
\begin{align*}
	&\pd{C}{t}= \od{C}{z}\pd{z}{t} = -\frac{z}{n t} \od{C}{z} 
	\\
	&\pd{C}{x} = \od{C}{z}\pd{z}{x} = \frac{1}{\lambda t^{1/n}} \od{C}{z} 
	\\
	&\pdd{C}{x} = \odd{C}{z} \left(\pd{z}{x}\right)^2 = \left(\frac{1}
		{\lambda t^{1/n}}\right)^2 \odd{C}{z}
\end{align*}

We shall leave the time derivative as it is and consider terms 
in Eq.~\ref{diff_eq_simpl_ap} involving spatial derivatives.
\begin{align*}
\pd{D}{x} \pd{C}{x} & = \frac{n k D_{GB}}{m(q \delta)^{m}x}\left(\frac{x}{\lambda}
\right)^{n}	\left[\frac{k}{(q\delta)^m}\left(t - \left(\frac{x}{\lambda}\right)^{n}
	\right)	\right]^{-\frac{1+m}{m}} \frac{1}{\lambda t^{1/n}} \od{C}{z}\\
	& = \od{C}{z}\cdot \frac{1}{\lambda t^{1/n}} \frac{n k D_{GB}}{m(q \delta)^{m}
	z \lambda t^{1/n}} z^n t \left[\frac{k}{(q\delta)^m}\left(t - z^n t\right)\right]
	^{-(1+m)/m} \\
	&=\od{C}{z}\frac{nq \delta D_{GB}}{mk^{1/m}\lambda^2}z^{n-1}\left(1 - z^n\right)^{-(1+m)/m}
	t^{-(\frac{1}{m} + \frac{2}{n})}
\end{align*}
\begin{align*}
	D\pdd{C}{x} & = D_{GB} \left[\frac{k}{(q\delta)^m}\left(t - \left(\frac{x}{\lambda}\right)^{n}
		\right)\right]^{-\frac{1}{m}} \odd{C}{z} \\
	& =\odd{C}{z} \cdot D_{GB} \frac{1}{\lambda^2 t^{2/n}}
	\left[\frac{k}{(q\delta)^m}\left(t - z^n t\right)\right]^{-1/m} \\
	& = \odd{C}{z} \frac{q \delta D_{GB}}{k^{1/m}\lambda^2}\left(1 - z^n\right)^{-1/m} t^{-(\frac{1}{m} + \frac{2}{n})}
\end{align*}

Note that $\pd{C}{t}$, $\pd{D}{x}\pd{C}{x}$ and $D\pdd{C}{x}$ depend only on $z$
and a power of time. For a solution to exist, the powers of time must cancel. 
It can be achieved with condition:
\begin{equation}
	\frac{1}{m} + \frac{2}{n} = 1
	\label{TheLink_ap}
\end{equation}
Thus, Eq.~\ref{TheLink_ap} connects the exponent of grain growth, $m$, and the 
exponent of the power law, $n$.

Now, we solve for $C(x, t)$. After the similarity transformation, 
Eq.~\ref{diff_eq_big} becomes:
\begin{equation}
	\odd{C}{z} \frac{q \delta D_{GB}}{k^{1/m}\lambda^2} \left(1 - z^n\right)^{-1/m}
	+ \od{C}{z} \left[ \frac{n}{m} \frac{q \delta D_{GB}}{k^{1/m}\lambda^2}
	z^{n-1}\left(1 - z^n\right)^{-(1+m)/m} + \frac{z}{n}\right] = 0 
\end{equation}

We introduce a dimensionless constant $\gamma = \lambda k^{1/2m} / \sqrt{q \delta D_{GB}}$
and use Eq.~\ref{TheLink_ap}, changing $1/m$ to $(1 - 2/n)$ and $n/m$ to $(n-2)$:
\begin{equation}
	\odd{C}{z} \frac{1}{\gamma^2} \left(1 - z^n\right)^{2/n - 1} 
	+\od{C}{z} \left[ \frac{n-2}{\gamma^2} z^{n-1}\left(1 - z^n\right)^{2/n - 2}
	+ \frac{z}{n}\right] = 0 
	\label{eq:eq_cond_ap}
\end{equation}

Integration of Eq.~\ref{eq:eq_cond_ap} gives the following solution:
\begin{equation*}
	C(x, t) = A_1 - A_2\bigintssss_{0}^{x/\lambda t^{1/n}} {(1 - y^n)^{1 - 2/n} 
	\exp\left[-\frac{\gamma^2}{2n}y^2\,
F\left(\frac{2 - n}{n}, \frac{2}{n}; \frac{2+n}{n}; y^n\right)\right]\,\mathrm{d}y}
\end{equation*}
where $A_1$ and $A_2$ are constants of integration, $F(\frac{2 - n}{n}, \frac{2}{n};
\frac{2+n}{n}; y^n)$ is a Gauss hypergeometric series.

Condition (Eq.~\ref{eq:bnd_cnd_ap}) on the \textit{A/PL} boundary gives
$A_1 = C_{max}$, while condition on the \textit{PL/B} boundary gives:
\begin{equation*}
	A_2 = (C_{max} - C_{eq}) \Big/ \bigintssss_{0}^{1} {(1 - y^n)^{1 - 2/n} 
	\exp\left[-\frac{\gamma^2}{2n}y^2\,
F\left(\frac{2- n}{n}, \frac{2}{n}; \frac{2+n}{n}; y^n\right)\right]\,\mathrm{d}y}
\end{equation*}

The value of yet undetermined constant $\gamma$ can be obtained from the condition 
on the moving boundary (Eq.~\ref{eq:stef_bc_ap}). This condition produces a transcendental
equation on $\gamma$, so it must be sought numerically in every particular
case. Thus, a solution for a solid-state reaction, limited by diffusion, with
grain-growth of the \textit{PL}, is given by:
\begin{numcases}{}
	 \ell(t) = \gamma\sqrt{\frac{q\delta D_{GB}}{k^{1/m}}}\, t^{1/n}\\
	C(x, t) = C_{max} - (C_{max} - C_{eq}) \frac{\int_{0}^{x/\lambda t^{1/n}}
		{(1 - y^n)^{1 - 2/n} \exp\left[-\frac{\gamma^2}{2n}y^2\,
F\left(\frac{2 - n}{n}, \frac{2}{n}; \frac{2+n}{n}; y^n\right)\right]\,\mathrm{d}y}}
{\int_{0}^{1} {(1 - y^n)^{1 - 2/n} 
	\exp\left[-\frac{\gamma^2}{2n}y^2\,
F\left(\frac{2- n}{n}, \frac{2}{n}; \frac{2+n}{n}; y^n\right)\right]\,\mathrm{d}y}}\\
  \begin{split}
	  n(C_{max} - C_{eq})\tilde{V}_A =& \gamma^2 \exp\left[-\frac{\gamma^2}{2n}
    \,F\left(\frac{2- n}{n}, \frac{2}{n}; \frac{2+n}{n}; 1\right)\right] \\
	&\times \bigintssss_{0}^{1} {(1 - y^n)^{1 - 2/n}\exp\left[-\frac{\gamma^2}{2n}y^2\,
    F\left(\frac{2 - n}{n}, \frac{2}{n}; \frac{2+n}{n}; y^n\right)\right]
\,\mathrm{d}y}
  \end{split}
\end{numcases}


\begin{thebibliography}{10}
	\expandafter\ifx\csname url\endcsname\relax
	\def\url#1{\texttt{#1}}\fi
	\expandafter\ifx\csname urlprefix\endcsname\relax\def\urlprefix{URL }\fi
	\expandafter\ifx\csname href\endcsname\relax
	\def\href#1#2{#2} \def\path#1{#1}\fi
	
	\bibitem{Wu2017_1}
	W.-P. Wu, Z.-F. Chen, Iridium coating: Processes, properties and application. Part I, Johnson Matthey Technology Review 61 (2017) 16--28.
	\newblock \href {https://doi.org/10.1595/205651317x693606}
	{\path{doi:10.1595/205651317x693606}}.
	
	\bibitem{Wu2017_2}
	W.-P. Wu, Z.-F. Chen, Iridium coating: Processes, properties and
	application. Part II, Johnson Matthey Technology Review 61 (2017) 93--110.
	\newblock \href {https://doi.org/10.1595/205651317X695064}
	{\path{doi:10.1595/205651317X695064}}.
	
	\bibitem{Arblaster1995}
	J.~Arblaster, The thermodynamic properties of iridium on ITS-90, Calphad 19
	(1995) 365--372.
	\newblock \href {https://doi.org/10.1016/0364-5916(95)00034-c}
	{\path{doi:10.1016/0364-5916(95)00034-C}}.
	
	\bibitem{Criscione1964}
	J.~M. Criscione, R.~A. Mercuri, E.~P. Schram, A.~W. Smith, H.~F. Volk, High
	temperature protective coatings for graphite --- Part II, Tech. rep., Air Force
	Materials Laboratory (1964).
	
	\bibitem{Mumtaz1995}
	K.~Mumtaz, J.~Echigoya, H.~Enoki, T.~Hirai, Y.~Shindo, Thermal cycling of
	iridium coatings on isotropic graphite, J. Mater. Sci. 30
	(1995) 465--472.
	\newblock \href {https://doi.org/10.1007/bf00354413}
	{\path{doi:10.1007/bf00354413}}.
	
	\bibitem{Wu2011}
	W.~Wu, Z.~Chen, H.~Cheng, L.~Wang, Y.~Zhang, Tungsten and iridium multilayered
	structure by DGP as ablation-resistance coatings for graphite, Appl.
	Surf. Sci. 257 (2011) 7295--7304.
	\newblock \href {https://doi.org/10.1016/j.apsusc.2011.03.108}
	{\path{doi:10.1016/j.apsusc.2011.03.108}}.
	
	\bibitem{Zhu2013}
	L.~Zhu, S.~Bai, H.~Zhang, Y.~Ye, W.~Gao, Rhenium used as an interlayer between
	carbon–carbon composites and iridium coating: Adhesion and wettability,
	Surf. Coat. Technol. 235 (2013) 68--74.
	\newblock \href {https://doi.org/10.1016/j.surfcoat.2013.07.013}
	{\path{doi:10.1016/j.surfcoat.2013.07.013}}.
	
	\bibitem{Zhu2014}
	L.~Zhu, S.~Bai, H.~Zhang, Y.~Ye, W.~Gao, Double-layer iridium–aluminum
	intermetallic coating on iridium/rhenium coated graphite prepared by pack
	cementation, Surf. Coat. Technol. 258 (2014) 524--530.
	\newblock \href {https://doi.org/10.1016/j.surfcoat.2014.08.044}
	{\path{doi:10.1016/j.surfcoat.2014.08.044}}.
	
	\bibitem{Chen2014}
	Z.~Chen, W.~Wu, X.~Cong, Oxidation resistance coatings of Ir–Zr and Ir by
	double glow plasma, J. Mater. Sci. Technol. 30 (2014)
	268--274.
	\newblock \href {https://doi.org/10.1016/j.jmst.2013.06.002}
	{\path{doi:10.1016/j.jmst.2013.06.002}}.
	
	\bibitem{Zhu2017}
	L.~Zhu, G.~Du, S.~Bai, H.~Zhang, Y.~Ye, Y.~Ai, Oxidation behavior of a
	double-layer iridium-aluminum intermetallic coating on iridium at the
	temperature of 1400°C–2000°C in the air atmosphere, Corros. Sci. 123
	(2017) 328--338.
	\newblock \href {https://doi.org/10.1016/j.corsci.2017.05.010}
	{\path{doi:10.1016/j.corsci.2017.05.010}}.
	
	\bibitem{Zhang2020}
	K.~Zhang, L.~Zhu, S.~Bai, Y.~Ye, H.~Zhang, S.~Li, Y.~Tang, J.~Zhang, G.~Wang,
	Ablation behavior of an Ir-Hf coating: A novel idea for ultra-high
	temperature coatings in non-equilibrium conditions, J. Alloy. Compd. 818 (2020) 152829.
	\newblock \href {https://doi.org/10.1016/j.jallcom.2019.152829}
	{\path{doi:10.1016/j.jallcom.2019.152829}}.
	
	\bibitem{Zhang2022}
	H.~Zhang, L.-A. Zhu, S.-X. Bai, Y.-C. Ye, Ablation-resistant Ir/Re coating on
	C/C composites at ultra-high temperatures, Rare Metals 41 (2022) 199--208.
	\newblock \href {https://doi.org/10.1007/s12598-015-0509-2}
	{\path{doi:10.1007/s12598-015-0509-2}}.
	
	\bibitem{Li2024}
	F.~Li, L.~Zhu, K.~Zhang, Z.~Wang, Y.~Ye, S.~Li, Y.~Tang, S.~Bai, Ir–Zr
	coating through a pack cementation method for ultra-high temperature
	applications, J. Mater. Res. Technol. 28 (2024)
	4428--4436.
	\newblock \href {https://doi.org/10.1016/j.jmrt.2024.01.067}
	{\path{doi:10.1016/j.jmrt.2024.01.067}}.
	
	\bibitem{Nikiforov2024}
	Y.~A. Nikiforov, V.~A. Danilovsky, V.~V. Lozanov, N.~I. Baklanova,
	High-temperature solid-state reaction between zirconium carbide and iridium:
	New insights into the phase formation, J. Am. Ceram.
	Soc. (2024).
	\newblock \href {https://doi.org/10.1111/jace.19675}
	{\path{doi:10.1111/jace.19675}}.
	
	\bibitem{Kwon1989}
	J.-W. Kwon, Formation and growth of \ce{Ir_3Hf} layers at Ir/HfC interfaces between
	1900°C and 2200°C (1989) 133.
	
	\bibitem{Park1999}
	J.~Park, K.~Landry, J.~Perepezko, Kinetic control of silicon carbide/metal
	reactions, Mater. Sci. Eng.: A 259 (1999) 279--286.
	\newblock \href {https://doi.org/10.1016/s0921-5093(98)00899-5}
	{\path{doi:10.1016/s0921-5093(98)00899-5}}.
	
	\bibitem{Bhanumurthy2001}
	K.~Bhanumurthy, R.~Schmid-Fetzer, Interface reactions between silicon carbide
	and metals (Ni, Cr, Pd, Zr), Composites Part A: Appl. Sci.
	Manuf. 32 (2001) 569--574.
	\newblock \href {https://doi.org/10.1016/s1359-835x(00)00049-x}
	{\path{doi:10.1016/s1359-835x(00)00049-x}}.
	
	\bibitem{Demkowicz2008}
	P.~Demkowicz, K.~Wright, J.~Gan, D.~Petti, High temperature interface reactions
	of TiC, TiN, and SiC with palladium and rhodium, Solid State Ionics 179
	(2008) 2313--2321.
	\newblock \href {https://doi.org/10.1016/j.ssi.2008.07.021}
	{\path{doi:10.1016/j.ssi.2008.07.021}}.
	
	\bibitem{Lopez2010}
	E.~López‐Honorato, K.~Fu, P.~J. Meadows, J.~Tan, P.~Xiao, Effect of
	microstructure on the resilience of silicon carbide to palladium attack,
	J. Am. Ceram. Soc. 93 (2010) 4135--4141.
	\newblock \href {https://doi.org/10.1111/j.1551-2916.2010.04005.x}
	{\path{doi:10.1111/j.1551-2916.2010.04005.x}}.
	
	\bibitem{Gentile2015}
	M.~Gentile, P.~Xiao, T.~Abram, Palladium interaction with silicon carbide,
	J. Nucl. Mater. 462 (2015) 100--107.
	\newblock \href {https://doi.org/10.1016/j.jnucmat.2015.03.013}
	{\path{doi:10.1016/j.jnucmat.2015.03.013}}.
	
	\bibitem{Golosov2023}
	M.~A. Golosov, A.~V. Utkin, V.~V. Lozanov, A.~T. Titov, N.~I. Baklanova,
	Microstructural patterning of the reaction zone formed by solid-state
	interaction between iridium and sic ceramics, Materialia 27 (3 2023).
	\newblock \href {https://doi.org/10.1016/j.mtla.2022.101647}
	{\path{doi:10.1016/j.mtla.2022.101647}}.
	
	\bibitem{Harrison1969}
	L.~Harrison, The theory of solid phase kinetics, in: C.~Bamford, C.~Tiper
	(Eds.), The Theory of Kinetics, Vol.~2 of Comprehensive Chemical Kinetics,
	Elsevier, 1969, pp. 377--462.
	\newblock \href {https://doi.org/10.1016/b978-0-444-40674-3.50011-0}
	{\path{doi:10.1016/b978-0-444-40674-3.50011-0}}.
	
	\bibitem{Suzuki2005}
	K.~Suzuki, S.~Kano, M.~Kajihara, N.~Kurokawa, K.~Sakamoto, Reactive diffusion
	between Ag and Sn at solid state temperatures, Materials Transactions 46
	(2005) 969--973.
	\newblock \href {https://doi.org/10.2320/matertrans.46.969}
	{\path{doi:10.2320/matertrans.46.969}}.
	
	\bibitem{Xu2006}
	L.~Xu, Y.~Y. Cui, Y.~L. Hao, R.~Yang, Growth of intermetallic layer in
	multi-laminated Ti/Al diffusion couples, Mater. Sci. Eng.: A
	435-436 (2006) 638--647.
	\newblock \href {https://doi.org/10.1016/j.msea.2006.07.077}
	{\path{doi:10.1016/j.msea.2006.07.077}}.
	
	\bibitem{Ren2013}
	Z.~Ren, X.~Hu, X.~Xue, K.~Chou, Solid state reaction studies in \ce{Fe_3O_4}–\ce{TiO_2}
	system by diffusion couple method, J. Alloy. Compd. 580 (2013)
	182--186.
	\newblock \href {https://doi.org/10.1016/j.jallcom.2013.05.114}
	{\path{doi:10.1016/j.jallcom.2013.05.114}}.
	
	\bibitem{Gueydan2014}
	A.~Gueydan, B.~Domengès, E.~Hug, Study of the intermetallic growth in
	copper-clad aluminum wires after thermal aging, Intermetallics 50 (2014)
	34--42.
	\newblock \href {https://doi.org/10.1016/j.intermet.2014.02.007}
	{\path{doi:10.1016/j.intermet.2014.02.007}}.
	
	\bibitem{Vianco1994}
	P.~T. Vianco, K.~L. Erickson, P.~L. Hopkins, Solid state intermetallic compound
	growth between copper and high temperature, tin-rich solders — Part I:
	Experimental analysis, J. Electron. Mater. 23 (1994) 721--727.
	\newblock \href {https://doi.org/10.1007/bf02651365}
	{\path{doi:10.1007/bf02651365}}.
	
	\bibitem{Bader1995}
	S.~Bader, W.~Gust, H.~Hieber, Rapid formation of intermetallic compounds
	interdiffusion in the Cu-Sn and Ni-Sn systems, Acta Metallurgica et
	Materialia 43 (1995) 329--337.
	\newblock \href {https://doi.org/10.1016/0956-7151(95)90289-9}
	{\path{doi:10.1016/0956-7151(95)90289-9}}.
	
	\bibitem{Mita2005}
	M.~Mita, M.~Kajihara, N.~Kurokawa, K.~Sakamoto, Growth behavior of \ce{Ni_3Sn_4} layer
	during reactive diffusion between Ni and Sn at solid-state temperatures,
	Mater. Sci. Eng.: A 403 (2005) 269--275.
	\newblock \href {https://doi.org/10.1016/j.msea.2005.05.012}
	{\path{doi:10.1016/j.msea.2005.05.012}}.
	
	\bibitem{Sakama2009}
	T.~Sakama, M.~Kajihara, Influence of Ag on kinetics of solid-state reactive
	diffusion between Pd and Sn, Materials Transactions 50 (2009) 266--274.
	\newblock \href {https://doi.org/10.2320/matertrans.mra2008292}
	{\path{doi:10.2320/matertrans.mra2008292}}.
	
	\bibitem{Li2010}
	J.~Li, P.~Agyakwa, C.~Johnson, Kinetics of \ce{Ag_3Sn} growth in Ag–Sn–Ag system
	during transient liquid phase soldering process, Acta Materialia 58 (2010)
	3429--3443.
	\newblock \href {https://doi.org/10.1016/j.actamat.2010.02.018}
	{\path{doi:10.1016/j.actamat.2010.02.018}}.
	
	\bibitem{Mirjalili2013}
	M.~Mirjalili, M.~Soltanieh, K.~Matsuura, M.~Ohno, On the kinetics of \ce{TiAl_3}
	intermetallic layer formation in the titanium and aluminum diffusion couple,
	Intermetallics 32 (2013) 297--302.
	\newblock \href {https://doi.org/10.1016/j.intermet.2012.08.017}
	{\path{doi:10.1016/j.intermet.2012.08.017}}.
	
	\bibitem{Lis2014}
	A.~Lis, M.~S. Park, R.~Arroyave, C.~Leinenbach, Early stage growth
	characteristics of \ce{Ag_3Sn} intermetallic compounds during solid-solid and
	solid-liquid reactions in the Ag-Sn interlayer system: Experiments and
	simulations, J. Alloy. Compd. 617 (2014) 763--773.
	\newblock \href {https://doi.org/10.1016/j.jallcom.2014.08.082}
	{\path{doi:10.1016/j.jallcom.2014.08.082}}.
	
	\bibitem{Zhang2019}
	C.~Q. Zhang, W.~Liu, Abnormal effect of temperature on intermetallic compound
	layer growth at aluminum-titanium interface: The role of grain boundary
	diffusion, Mater. Lett. 254 (2019) 1--4.
	\newblock \href {https://doi.org/10.1016/j.matlet.2019.07.013}
	{\path{doi:10.1016/j.matlet.2019.07.013}}.
	
	
	\bibitem{Oh2024}
	M.~Oh, K.~Matsushita, E.~Kobayashi, M.~Kajihara, The growth kinetics of intermetallic compounds
	by the fast diffusion path at the interface of Co and molten Zn,
	J. Mol. Liq. 413 (2024) 125966.
	\newblock \href{https://doi.org/10.1016/j.molliq.2024.125966}{\path{doi:10.1016/j.molliq.2024.125966}}.
	
	\bibitem{Murray1988}
	P.~Murray, G.~Carey, Finite element analysis of diffusion with reaction at a
	moving boundary, J. Comput. Phys. 74 (1988) 440--455.
	\newblock \href {https://doi.org/10.1016/0021-9991(88)90087-3}
	{\path{doi:10.1016/0021-9991(88)90087-3}}.
	
	\bibitem{Wagner1969}
	C.~Wagner, The evaluation of data obtained with diffusion couples of binary
	single-phase and multiphase systems, Acta Metallurgica 17 (1969) 99--107.
	\newblock \href {https://doi.org/10.1016/0001-6160(69)90131-x}
	{\path{doi:10.1016/0001-6160(69)90131-x}}.
	
	\bibitem{Lichtner1986}
	P.~C. Lichtner, E.~H. Oelkers, H.~C. Helgeson, Exact and numerical solutions to
	the moving boundary problem resulting from reversible heterogeneous reactions
	and aqueous diffusion in a porous medium, J. Geophys. Res.:
	Solid Earth 91 (1986) 7531--7544.
	\newblock \href {https://doi.org/10.1029/jb091ib07p07531}
	{\path{doi:10.1029/jb091ib07p07531}}.
	
	\bibitem{Erickson1994}
	K.~L. Erickson, P.~L. Hopkins, P.~Vianco, Solid state intermetallic compound
	growth between copper and high temperature, tin-rich solders — Part II:
	Modeling, J. Electron. Mater. 23 (1994) 729--734.
	\newblock \href {https://doi.org/10.1007/bf02651366}
	{\path{doi:10.1007/bf02651366}}.
	
	\bibitem{Schaefer1998}
	M.~Schaefer, R.~A. Fournelle, J.~Liang, Theory for intermetallic phase growth
	between cu and liquid Sn-Pb solder based on grain boundary diffusion control,
	J. Electron. Mater. 27 (1998) 1167--1176.
	\newblock \href {https://doi.org/10.1007/s11664-998-0066-7}
	{\path{doi:10.1007/s11664-998-0066-7}}.
	
	\bibitem{Ghosh2000}
	G.~Ghosh, Coarsening kinetics of \ce{Ni_3Sn_4} scallops during interfacial reaction
	between liquid eutectic solders and Cu/Ni/Pd metallization, J.
	Appl. Phys. 88 (2000) 6887--6896.
	\newblock \href {https://doi.org/10.1063/1.1321791}
	{\path{doi:10.1063/1.1321791}}.
	
	\bibitem{Wang2015}
	L.~Wang, Y.~Wang, P.~Prangnell, J.~Robson, Modeling of intermetallic compounds
	growth between dissimilar metals, Metall. Mater. Trans. A:
	Phys. Metall. Mater. Sci. 46 (2015) 4106--4114.
	\newblock \href {https://doi.org/10.1007/s11661-015-3037-7}
	{\path{doi:10.1007/s11661-015-3037-7}}.
	
	\bibitem{Xu2018}
	L.~Xu, J.~D. Robson, L.~Wang, P.~B. Prangnell, The influence of grain structure
	on intermetallic compound layer growth rates in Fe-Al dissimilar welds,
	Metall. Mater. Trans. A: Phys. Metall. Mater. Sci. 49 (2018) 515--526.
	\newblock \href {https://doi.org/10.1007/s11661-017-4352-y}
	{\path{doi:10.1007/s11661-017-4352-y}}.
	
	\bibitem{Mishin1997}
	Y.~Mishin, C.~Herzig, J.~Bernardini, W.~Gust, Grain boundary diffusion:
	Fundamentals to recent developments, Int. Mater. Rev. 42
	(1997) 155--178.
	\newblock \href {https://doi.org/10.1179/imr.1997.42.4.155}
	{\path{doi:10.1179/imr.1997.42.4.155}}.
	
	\bibitem{Abbaschian2008}
	R.~Abbaschian, L.~Abbaschian, R.~E. Reed-Hill, {Physical Metallurgy
		Principles}, 4th Edition, Cengage Learning, 2008.
	
	\bibitem{Molodov2013}
	D.~A. Molodov, {Microstructural Design of Advanced Engineering Materials},
	2013.
	\newblock \href {https://doi.org/10.1002/9783527652815}
	{\path{doi:10.1002/9783527652815}}.
	
	\bibitem{Hu1970}
	H.~Hu, B.~B. Rath, On the time exponent in isothermal grain growth,
	Metall. Trans. 1 (1970) 3181--3184.
	\newblock \href {https://doi.org/10.1007/bf03038435}
	{\path{doi:10.1007/bf03038435}}.
	
	\bibitem{Louat1974}
	N.~P. Louat, {On the theory of normal grain growth}, Acta Metallurgica 22~(6)
	(1974) 721--724.
	\newblock \href {https://doi.org/10.1016/0001-6160(74)90081-9}
	{\path{doi:10.1016/0001-6160(74)90081-9}}.
	
	\bibitem{Anderson1989}
	M.~P. Anderson, G.~S. Grest, D.~J. Srolovitz, {Computer simulation of normal
		grain growth in three dimensions}, Philos. Mag. B 59~(3) (1989)
	293--329.
	\newblock \href {https://doi.org/10.1080/13642818908220181}
	{\path{doi:10.1080/13642818908220181}}.
	
	\bibitem{Gao1996}
	J.~Gao, R.~G. Thompson, {Real time-temperature models for Monte Carlo
		simulations of normal grain growth}, Acta Materialia 44~(11) (1996)
	4565--4570.
	\newblock \href {https://doi.org/10.1016/1359-6454(96)00079-1}
	{\path{doi:10.1016/1359-6454(96)00079-1}}.
	
	\bibitem{Liu2001}
	G.~Liu, H.~Yu, X.~Song, X.~Qin, {A new model of three-dimensional grain growth:
		Theory and computer simulation of topology-dependency of individual grain
		growth rate}, Mater. Des. 22~(1) (2001) 33--38.
	\newblock \href {https://doi.org/10.1016/s0261-3069(00)00040-6}
	{\path{doi:10.1016/s0261-3069(00)00040-6}}.
	
	\bibitem{Yu2003}
	Q.~Yu, S.~K. Esche, {Three-dimensional grain growth modeling with a Monte Carlo
		algorithm}, Mater. Lett. 57~(30) (2003) 4622--4626.
	\newblock \href {https://doi.org/10.1016/s0167-577x(03)00372-0}
	{\path{doi:10.1016/s0167-577x(03)00372-0}}.
	
	\bibitem{Najafkhani2021}
	F.~Najafkhani, S.~Kheiri, B.~Pourbahari, H.~Mirzadeh, Recent advances in the
	kinetics of normal/abnormal grain growth: a review, Arch. Civ. Mech. Eng. 21 (2021) 29.
	\newblock \href {https://doi.org/10.1007/s43452-021-00185-8}
	{\path{doi:10.1007/s43452-021-00185-8}}.
	
	\bibitem{AbramowitzStegun}
	F.~Oberhettinger, Hypergeometric functions, in: M.~Abramowitz, I.~Stegun
	(Eds.), Handbook of Mathematical Functions, 10th Edition, National Bureau of
	Standards, 1972, Ch.~15, pp. 555--566.
	
	\bibitem{Schneider2012}
	C.~A. Schneider, W.~S. Rasband, K.~W. Eliceiri, NIH Image to ImageJ: 25 years
	of image analysis, Nature Methods 2012 9:7 9 (2012) 671--675.
	\newblock \href {https://doi.org/10.1038/nmeth.2089}
	{\path{doi:10.1038/nmeth.2089}}.
	
	\bibitem{Ran2006}
	H.~Ran, Z.~Du, Thermodynamic assessment of the Ir–Zr system, J.
	Alloy. Compd. 413 (2006) 101--105.
	\newblock \href {https://doi.org/10.1016/j.jallcom.2005.06.060}
	{\path{doi:10.1016/j.jallcom.2005.06.060}}.
	
	\bibitem{Brewer1973}
	L.~Brewer, P.~Wengert, Thermodynamic stability of certain intermetallic
	compounds made from transition elements, Metall. Trans. (1973) 15--18.
	
	\bibitem{Ikeda1998}
	T.~Ikeda, A.~Almazouzi, H.~Numakura, M.~Koiwa, W.~Sprengel, H.~Nakajima,
	Single-phase interdiffusion in \ce{Ni_3Al}, Acta Materialia 46 (1998) 5369--5376.
	\newblock \href {https://doi.org/10.1016/s1359-6454(98)00209-2}
	{\path{doi:10.1016/s1359-6454(98)00209-2}}.
	
	\bibitem{Numakura2001}
	H.~Numakura, T.~Ikeda, H.~Nakajima, M.~Koiwa, Diffusion in \ce{Ni_3Al}, \ce{Ni_3Ga} and \ce{Ni_3Ge},
	Mater. Sci. Eng.: A 312 (2001) 109--117.
	\newblock \href {https://doi.org/10.1016/s0921-5093(00)01864-5}
	{\path{doi:10.1016/s0921-5093(00)01864-5}}.
	
	\bibitem{Uchida2005}
	M.~Uchida, H.~Numakura, Y.~Yamabe-Mitarai, E.~Bannai, Chemical diffusion in
	\ce{Ir_3Nb}, Scr. Materialia 52 (2005) 11--15.
	\newblock \href {https://doi.org/10.1016/j.scriptamat.2004.09.006}
	{\path{doi:10.1016/j.scriptamat.2004.09.006}}.
	
	\bibitem{Mehrer1990}
	H.~Mehrer, N.~Stolica, N.~A. Stolwijk, 2.2.10 Nickel group metals, Vol.~26,
	Springer-Verlag, 1990, pp. 52--54.
	\newblock \href {https://doi.org/10.1007/b37801} {\path{doi:10.1007/b37801}}.
	
	\bibitem{Grey1973}
	E.~A. Grey, G.~T. Higgins, Solute limited grain boundary migration: A
	rationalisation of grain growth, Acta Metall. 21 (1973) 309--321.
	\newblock \href {https://doi.org/10.1016/0001-6160(73)90186-7}
	{\path{doi:10.1016/0001-6160(73)90186-7}}.
	
	\bibitem{Kim2009}
	H.~S. Kim, H.~J. Lee, Y.~S. Yu, Y.~S. Won, Three-dimensional simulation of
	intermetallic compound layer growth in a binary alloy system, Acta Materialia
	57 (2009) 1254--1262.
	\newblock \href {https://doi.org/10.1016/j.actamat.2008.11.006}
	{\path{doi:10.1016/j.actamat.2008.11.006}}.
	
\end{thebibliography}

\end{document}